\documentclass[sn-mathphys,Numbered]{sn-jnl}

\usepackage{graphicx}
\usepackage{multirow}
\usepackage{amsmath,amssymb,amsfonts}
\usepackage{amsthm}
\usepackage{mathrsfs}
\usepackage[title]{appendix}
\usepackage{xcolor}
\usepackage{textcomp}
\usepackage{manyfoot}
\usepackage{booktabs}
\usepackage{algorithm}
\usepackage{algorithmicx}
\usepackage{algpseudocode}
\usepackage{listings}
\usepackage[utf8]{inputenc}
\usepackage{hyperref}
\usepackage{verbatim}

\newcommand*{\Apart}{\ensuremath{\langle A_\mathrm{part} \rangle_\mathrm{b}} }

\newcommand*{\sqSnn}{\ensuremath{\sqrt{s_\mathrm{NN}}} }

\newcommand*{\Tbeam}{\ensuremath{T_\mathrm{b}} }

\begin{document}

\title[Article Title]{Systematics of yields of strange hadrons from heavy-ion collisions around threshold energies}

\author[1]{\fnm{K.} \sur{Piasecki}}

\author[2]{\fnm{P.} \sur{Piotrowski}} 

\affil[1]{\orgname{University of Warsaw}, \orgaddress{\street{Pasteura 5}, \city{Warsaw}, \postcode{02-093}, \country{Poland}}}

\abstract{
The parametrizations of experimental yields of K$^{\pm,0}$, 
$\phi$ and $\Lambda+\Sigma^0$ are proposed as function of 
available energy, \sqSnn, and number of participants, \Apart,
for \sqSnn from 2.15 to 3 GeV. For all the dataset 
the \Apart was extracted using the Glauber Monte Carlo method. 
The $\alpha$ exponent of yield dependency on \Apart appears 
not to change with beam energy and is found to be
\mbox{1.30 $\pm$ 0.02}. Our parametrization and the predictions of
public versions of RQMD.RMF, SMASH and UrQMD transport models 
are compared to the HADES experimental data for Ar+KCl at \sqSnn of 2.61 GeV. 
The phenomenological parametrization currently offers 
the best overall description of these yields. 
Predictions are given for yields from Ag+Ag collisions 
at available energies of 2.41 and 2.55 GeV, analysed by HADES, 
Au+Au experiment at 2.16 and 2.24 GeV planned by this collaboration,
some unpublished yields for STAR's Au+Au collisions at 3 GeV,
and for Au+Au collisions planned by CBM, up to 3.85 GeV.
}
\keywords{strange hadron yields, relativistic heavy ion collisions}



\maketitle

\section{Introduction}

Three decades after the first systematics of meson production at 
beam energies around their thresholds in the free nucleon-nucleon (NN) 
channel \cite{Metag93}, a considerable sample of yields of hadrons
with strange valence quarks was obtained with help of the KaoS, FOPI 
and HADES setups installed at the SIS18 accelerator in the GSI.
STAR recently explored this region with measurements at the available 
energy in the NN channel of $\sqSnn = 3$~GeV~\cite{Abdallah22}
\footnote{For consistency, all the energies of colliding systems
were given in terms of available energy in the nucleon-nucleon channel.
For orientation, the beam kinetic energies per nucleon are shown 
in relevant tables.}.
For HADES, the analysis of strangeness from Ag+Ag at available energies
of 2.41 and 2.55 GeV is ongoing~\cite{Spies22} and the new experiment 
at energies between 1.95 and 2.25 GeV is planned. 
Also, the CBM being constructed together with SIS100 accelerator 
is expected to deliver soon new strangeness data from heavy-ion 
collisions at a few GeV~\cite{Agarwal23}. 
With this in mind, it is a good time to compile the state-of-art 
systematics of yields ($P$) of hadrons with strange quarks at 
energies around thresholds for strangeness production. 
In this paper we present such systematics for K$^{+,-,0}$, $\phi$ and 
$\Lambda + \Sigma^0$. However, we omit $\Xi^-$ and strange resonances 
due to currently too limited data to provide the systematics.
As for nearly all the heavy-ion experiments the $\Sigma^0$ decays 
into $\Lambda$ were indistinguishable from $\Lambda$ produced directly
in the collision zone, the yields of both hyperons were presented together.
Therefore, in our paper we follow this convention.

The values of threshold available energy needed for the production
of the considered hadrons in a free NN collision are:
2.55 GeV for K$^+$, K$^0$ and $\Lambda$ (within the NN $\rightarrow$ NK$\Lambda$ channel),
2.86 GeV for K$^-$ (produced in the NN $\rightarrow$ NNK$^+$K$^-$ process)
and 2.90 GeV for $\phi$ meson (for the NN $\rightarrow$ NN$\phi$ channel).
A sub-threshold production of these hadrons in the heavy-ion collisions
is possible due to opening up of many alternatives to direct NN collisions.
For example, calculations within the transport models in the considered
region of energies point to $\pi$Y and BY (B = baryon, Y = hyperon) 
-initiated channels~\cite{Schade10,Hartnack12} as possibly the leading sources 
of K$^-$ mesons. The experiments together with some of these models 
reveal also a considerable feeddown from $\phi$ meson decays~\cite{Stei16,Song22}.
In case of the latter hadrons, a broad mixture of channels may be
responsible for their production, possibly including feeddowns 
from decays of N$^*$ and $\Delta$ resonances~\cite{Schade10,Stei16,Song22}.

The obvious quantities on which the yields of considered hadrons 
should depend are \sqSnn and the average number of participant nucleons 
in the initial collision stage, \Apart,
the latter being the model-dependent function of the event centrality.
Such a systematics could be a handy tool in estimating and/or comparing 
the strangeness production for any analysed or future heavy-ion experiment 
at comparable energies. Also, it could be helpful for theorists, 
as any model postulating a mixture of channels leading to production 
(and absorption) of strangeness in this domain, should reproduce 
the measured dependency.
In our considerations we shall not go much beyond \mbox{$\sqSnn \sim 3$~GeV}, 
as the particle production mechanism is thought to pass from hadron-like 
to soft string-like nature. Mixing data from both regions to obtain one 
systematics may lead to biased conclusions. 
The dependency of the yield on \Apart is universally found to have a
power character~\cite{Barth97,Foerster07,Adamczewski19}. 
With an accumulated set of yields it is interesting to check whether
or not the extracted exponent is the same for all the investigated hadrons, 
and whether $\alpha$ changes or not with \sqSnn. 

An attempt to directly set together the \Apart values from all 
the published data encounters the problem of different modelling 
of the collision centrality by various authors.
For all the FOPI and part of the KaoS analyses the Geometrical model 
\cite{GeomModel} was used, wherein the nuclear density profile was 
coarsely approximated by the sharp sphere.
Within some KaoS works, the optical Glauber approach \cite{GlauberOpt}
was employed \cite{Foerster07}, however,
with only one parametrization of the nuclear shape 
("two-parameter Fermi", 2pF). 
For the recent analyses of HADES \cite{HadesCentr} and STAR \cite{Abdallah22}
the Glauber Monte Carlo simulations \cite{Miller07} were performed. 
Therefore, the first task of our analysis is adjustment of the \Apart 
estimations to an equal footing by consistent application of the 
Glauber Monte Carlo approach to all the data. 
While we certainly don't claim that our approach is final, we perceive 
it as a step forward compared to the geometrical sharp-sphere approach 
and optical Monte Carlo with just 2pF profile with fixed parameters. 

The paper is organized as follows: 
in Sect.~\ref{Sect:Apart} the collected 4$\pi$ yields are presented 
with emphasis on the centrality models used in the original papers. 
We then describe our procedure of \Apart determination within the 
Glauber Monte Carlo approach.
In Sect.~\ref{Sect:paramcurve} the results of the best fits of the 
parametrization curve to the yields as function of \sqSnn and \Apart
are presented.
The predictions of the curve for considered hadrons are compared to 
the range of known yields, and shown for some colliding systems 
where data is being analysed or planned to be measured. 
We also discuss the behaviour of $\alpha$ exponent of the dependency
of yield on \Apart.
In Sect.~\ref{Sect:transport} we compare the predictive power of our
parametrization with results of some publicly available transport codes 
for the collisions of Ar+KCl at \sqSnn = 2.61 GeV. We also present
the predictions of our parametrization together with results of
transport codes for central collisions of Ag+Ag at \sqSnn of 2.41 and 2.55 GeV.
Section \ref{Sect:summary} summarizes the paper.
Some additional information is given in Appendices.

\section{Data selection and \Apart modelling}
\label{Sect:Apart}

For the sample of strange hadron yields subjected to our analysis we accepted 
all the published 4$\pi$ yields of K$^{\pm,0}$ and $\phi$ mesons, 
as well as $\Lambda + \Sigma^0$ baryons emitted from the nucleus-nucleus 
collisions at \sqSnn = 2.16 - 4.86 GeV,
where the centralities or multiplicities of charged particles (MUL) were given.
We also extracted the yields of K$^\pm$ from central collisions of Ni+Ni 
at 2.67 GeV from their rapidity distributions, as the acceptance was wide 
and the data profile sufficient enough for the systematic uncertainty of 
the yield to be rather small \cite{Menzel00}.
We did not include the yield of K$^+$ emitted from Ar+Ti at very low available
energy of 1.92 MeV due to narrow acceptance, large relative uncertainty 
and no control of centrality~\cite{Julien91}.
In total we collected 107 data points: 44 for K$^+$, 30 for K$^-$, 
11 for K$^0_s$, 13 for $\Lambda$ and 9 for $\phi$. 
They are listed in Table \ref{tab:yields_kp} for K$^+$,
Table \ref{tab:yields_kmphi} for K$^-$ and $\phi$ and 
Table \ref{tab:yields_k0Lam} for K$^0_s$ and $\Lambda$.
One can also find there the published centralities (based on cuts on MUL), 
the \Apart values stated in the respective papers, and the types of models 
originally used by the Authors. 
Apparently the Geometrical Model dominates with 61 points (all the FOPI and 
AGS data, 60\% of KaoS points, HADES data for Ar+KCl at 
\mbox{\sqSnn = 2.61} GeV and some others). 
The Glauber Optical approach was used 19 times, exclusively by KaoS.
The Monte Carlo version of the Glauber model was applied 22 times (Au+Au 
data from HADES at 2.41 GeV and STAR data for 3 GeV). 
For the remaining 5 points only the centralities were given, not the \Apart.
Merging data with \Apart obtained by different models weakens 
control over the field. 
In the prevailing Geometrical Model the nuclear shape profile was assumed 
to be of the sharp ball, 
which cannot be justified in light of measurements of the nuclear density profiles~\cite{Vries87} 
and deformation models~\cite{Moller95}.
A need to correct the situation seems to be evident.
Therefore, we performed an unified procedure of extraction of \Apart 
for all the data points, within the Glauber Monte Carlo approach.
To achieve this we used the TGlauberMC code~\cite{Alver08,Loizides15,Loizides18}. 

\begin{table}[!htb]
 \centering
 \begin{tabular}{|c|c|c|c|c|c|c|c|c|c|}
  \hline
Data &  {}  &\sqSnn &$\Tbeam/A$&  Centrality  &                {}                &\multicolumn{2}{c|}{As published}&      This work      &        {}         \\
point&System& [GeV] &  [GeV]   &     [\%]     &               Yield              &     \Apart      &    Model      &        \Apart       &       Ref.        \\
  \hline
  1 & Au+Au & 2.156 &  0.6     &   0   - 100   & $(7.3  \pm 1.6 ) \cdot 10^{-5}$ &  98.5           &    geom.      &  75.7   $\pm$ 3.6   & \cite{Foerster07} \\
  2 &  C+C  & 2.242 &  0.8     &   0   - 100   & $(1.75 \pm 0.32) \cdot 10^{-5}$ &   6             &    geom.      &   5.67  $\pm$ 0.11  & \cite{Foerster07} \\
  3 & Au+Au & 2.242 &  0.8     &   0   - 100   & $(1.47 \pm 0.27) \cdot 10^{-3}$ &  98.5           &    geom.      &   88.0  $\pm$ 2.5   & \cite{Foerster07} \\
  4 &  C+C  & 2.324 &  1.0     &   0   - 100   & $(8.0  \pm 1.1 ) \cdot 10^{-5}$ &   6             &    geom.      &   5.92  $\pm$ 0.11  & \cite{Foerster07} \\
  5 & Ni+Ni & 2.324 &  1.0     &   0   -  11.4 & $(3.22 \pm 0.56) \cdot 10^{-3}$ &  85.7           &    geom.      &  76.49  $\pm$ 0.81  & \cite{Barth97}    \\
  6 & Ni+Ni & 2.324 &  1.0     &  11.4 -  17.9 & $(1.79 \pm 0.33) \cdot 10^{-3}$ &  61.6           &    geom.      &  57.62  $\pm$ 0.82  & \cite{Barth97}    \\
  7 & Ni+Ni & 2.324 &  1.0     &  17.9 -  41.0 & $(7.40 \pm 1.35) \cdot 10^{-4}$ &  37.6           &    geom.      &  36.45  $\pm$ 0.46  & \cite{Barth97}    \\
  8 & Ni+Ni & 2.324 &  1.0     &  41.0 -  59.6 & $(2.30 \pm 0.45) \cdot 10^{-4}$ &  16.2           &    geom.      &  17.25  $\pm$ 0.37  & \cite{Barth97}    \\
  9 & Au+Au & 2.324 &  1.0     &   0   - 100   & $(4.5  \pm 0.7 ) \cdot 10^{-3}$ &  98.5           &    geom.      &   89.8  $\pm$ 2.6   & \cite{Foerster07} \\
 10 & Ni+Ni & 2.348 &  1.06    &   0   -  12.9 & $(3.30 \pm 0.83) \cdot 10^{-3}$ &  75             &    geom.      &  75.31	$\pm$ 0.75  & \cite{Best97}     \\
 11 & Au+Au & 2.378 &  1.135   &   0   - 100   & $(9.4  \pm 2.1 ) \cdot 10^{-3}$ &  98.5           &    geom.      &   90.9	$\pm$ 2.6   & \cite{Foerster07} \\
 12 &  C+C  & 2.403 &  1.2     &   0   - 100   & $(3.10 \pm 0.46) \cdot 10^{-4}$ &   6             &    geom.      &   6.07  $\pm$ 0.12  & \cite{Foerster07} \\
 13 & Au+Au & 2.415 &  1.23    &   0   -  10   & $(5.98 \pm 0.68) \cdot 10^{-2}$ & 303.0 $\pm$ 11.0&  Glau. MC     &  300.4	$\pm$ 4.0   & \cite{Adamczewski18} \\
 14 & Au+Au & 2.415 &  1.23    &  10   -  20   & $(3.39 \pm 0.38) \cdot 10^{-2}$ & 213.1 $\pm$ 11.1&  Glau. MC     &  212.3	$\pm$ 3.8   & \cite{Adamczewski18} \\
 15 & Au+Au & 2.415 &  1.23    &  20   -  30   & $(1.88 \pm 0.21) \cdot 10^{-2}$ & 149.8 $\pm$ 9.7 &  Glau. MC     &  149.3	$\pm$ 4.2   & \cite{Adamczewski18} \\
 16 & Au+Au & 2.415 &  1.23    &  30   -  40   & $(1.20 \pm 0.14) \cdot 10^{-2}$ & 103.1 $\pm$ 6.8 &  Glau. MC     &  102.2	$\pm$ 4.2   & \cite{Adamczewski18} \\
 17 & Ni+Ni & 2.499 &  1.45    &   0   -  12.9 & $(1.95 \pm 0.50) \cdot 10^{-2}$ &  75             &    geom.      &  78.05	$\pm$ 0.70  & \cite{Best97}       \\
 18 &  C+C  & 2.518 &  1.5     &   0   - 100   & $(1.30 \pm 0.16) \cdot 10^{-3}$ &   6             &    geom.      &   6.27	$\pm$ 0.11  & \cite{Foerster07}   \\
 19 & Ni+Ni & 2.518 &  1.5     &   0   -   4.4 & $(3.11 \pm 0.36) \cdot 10^{-2}$ & 101             &  Glau. Opt    &  89.71	$\pm$ 0.66  & \cite{Foerster07}   \\
 20 & Ni+Ni & 2.518 &  1.5     &   4.4 -  15   & $(2.80 \pm 0.31) \cdot 10^{-2}$ &  77             &  Glau. Opt    &  70.85	$\pm$ 0.62  & \cite{Foerster07}   \\
 21 & Ni+Ni & 2.518 &  1.5     &  15   -  26.5 & $(1.68 \pm 0.19) \cdot 10^{-2}$ &  52.8           &  Glau. Opt    &  50.31	$\pm$ 0.54  & \cite{Foerster07}   \\
 22 & Ni+Ni & 2.518 &  1.5     &  26.5 -  45.9 & $(8.2  \pm 1.0 ) \cdot 10^{-3}$ &  31             &  Glau. Opt    &  30.15	$\pm$ 0.45  & \cite{Foerster07}   \\
 23 & Ni+Ni & 2.518 &  1.5     &  45.9 - 100   & $(1.19 \pm 0.15) \cdot 10^{-3}$ &   7             &  Glau. Opt    &   8.18	$\pm$ 0.18  & \cite{Foerster07}   \\
 24 & Au+Au & 2.518 &  1.5     &   0   -   5.4 & $(1.58 \pm 0.17) \cdot 10^{-1}$ & 336.2           &  Glau. Opt    &  329.2	$\pm$ 3.6   & \cite{Foerster07}   \\
 25 & Au+Au & 2.518 &  1.5     &   5.4 -  18.1 & $(1.16 \pm 0.12) \cdot 10^{-1}$ & 252             &  Glau. Opt    &  241.2	$\pm$ 3.4   & \cite{Foerster07}   \\
 26 & Au+Au & 2.518 &  1.5     &  18.1 -  31.1 & $(6.06 \pm 0.73) \cdot 10^{-2}$ & 164.8           &  Glau. Opt    &  154.2	$\pm$ 3.8   & \cite{Foerster07}   \\
 27 & Au+Au & 2.518 &  1.5     &  31.1 -  52.3 & $(2.40 \pm 0.28) \cdot 10^{-2}$ &  88.2           &  Glau. Opt    &   80.7	$\pm$ 3.6   & \cite{Foerster07}   \\
 28 & Au+Au & 2.518 &  1.5     &  52.3 - 100   & $(3.28 \pm 0.70) \cdot 10^{-3}$ &  16             &  Glau. Opt    &   16.1	$\pm$ 1.2   & \cite{Foerster07}   \\
 29 &Ar+KCl & 2.611 &  1.756   &   0   -  35   & $(2.80 \pm 0.24) \cdot 10^{-2}$ &  38.5           &    geom.      &  37.17	$\pm$ 0.59  & \cite{Agakishiev09} \\
 30 &  C+C  & 2.627 &  1.8     &   0   - 100   & $(3.18 \pm 0.32) \cdot 10^{-3}$ &   6             &    geom.      &   6.40	$\pm$ 0.11  & \cite{Laue00}       \\
 31 & Ni+Ni & 2.627 &  1.8     &   0   -  11.4 & $(6.38 \pm 0.99) \cdot 10^{-2}$ &  85.7           &    geom.      &  81.52	$\pm$ 0.59  & \cite{Barth97}      \\
 32 & Ni+Ni & 2.627 &  1.8     &  11.4 -  17.9 & $(4.23 \pm 0.66) \cdot 10^{-2}$ &  61.3           &    geom.      &  61.64	$\pm$ 0.54  & \cite{Barth97}      \\
 33 & Ni+Ni & 2.627 &  1.8     &  17.9 -  41.0 & $(1.78 \pm 0.28) \cdot 10^{-2}$ &  37.3           &    geom.      &  38.77	$\pm$ 0.45  & \cite{Barth97}      \\
 34 & Ni+Ni & 2.627 &  1.8     &  41.0 -  59.6 & $(3.75 \pm 0.61) \cdot 10^{-3}$ &  15.4           &    geom.      &  18.11	$\pm$ 0.36  & \cite{Barth97}      \\
 35 & Al+Al & 2.666 &  1.91    &   0   -   8.6 & $(3.60 \pm 0.49) \cdot 10^{-2}$ &  42             &    geom.      &  36.78	$\pm$ 1.13  & \cite{Gasik16}      \\
 36 & Ni+Ni & 2.666 &  1.91    &   0   -  56   & $(3.60 \pm 0.05) \cdot 10^{-2}$ &  46.5           &    geom.      &  44.81	$\pm$ 0.54  & \cite{Piasecki19}   \\
 37 & Ni+Ni & 2.673 &  1.93    &   0   -  12.9 & $(8.3  \pm 2.2 ) \cdot 10^{-2}$ &  75             &    geom.      &   80.2	$\pm$ 0.7   & \cite{Best97}       \\
 38 & Ni+Ni & 2.673 &  1.93    &   0   -  21.4 & $(6.6  \pm 0.3 ) \cdot 10^{-2}$ &  72.7           &    geom.      &   70.9	$\pm$ 0.7   & \cite{Menzel00}     \\
 39 &  C+C  & 2.698 &  2.0     &   0   - 100   & $(5.3  \pm 0.58) \cdot 10^{-3}$ &   6             &    geom.      &   6.47	$\pm$ 0.11  & \cite{Foerster07}   \\
 40 & Au+Au & 4.859 &  10.7    &   0   -   5   & $ 24.2 \pm 0.9   $              & 354             &    geom.      &  341.5 	$\pm$ 3.8   & \cite{Ahle98}       \\
 41 & Au+Au & 4.859 &  10.7    &   5   -  12   & $ 19.7 \pm 0.6   $              & 312             &    geom.      &  278.2 	$\pm$ 3.9   & \cite{Ahle98}       \\
 42 & Au+Au & 4.859 &  10.7    &  12   -  23   & $ 13.3 \pm 0.4   $              & 248             &    geom.      &  205.7 	$\pm$ 4.2   & \cite{Ahle98}       \\
 43 & Au+Au & 4.859 &  10.7    &  23   -  39   & $  8.0 \pm 0.3   $              & 164             &    geom.      &  128.1 	$\pm$ 4.6   & \cite{Ahle98}       \\
 44 & Au+Au & 4.859 &  10.7    &  39   -  75   & $  2.0 \pm 0.3   $              &  62             &    geom.      &   44.4 	$\pm$ 2.8   & \cite{Ahle98}       \\
  \hline
 \end{tabular}
 \caption{Yields of K$^+$ mesons emitted from nuclei colliding at given available energies (as well as beam kinetic energies) and centrality ranges. 
          Values of \Apart are shown as published originally and according to our Glauber Monte Carlo analysis (see text for details.)}
 \label{tab:yields_kp}
\end{table}

\begin{table}[hb]
 \setlength{\tabcolsep}{4.5pt}
 \centering
 \begin{tabular}{|c|c|c|c|c|c|c|c|c|c|}
  \hline
  {}   &\sqSnn &$\Tbeam/A$&   Centrality  &                    \multicolumn{2}{c|}{Yields}                    & \multicolumn{2}{c|}{As published}&    This work      &          {}        \\
System & [GeV] & [GeV]    &      [\%]     &             K$^-$              &               $\phi$             &     \Apart        &     Model    &     \Apart        &         Ref.       \\
  \hline
 Au+Au & 2.415 &  1.23    &   0   -  20   & $(3.36 \pm 0.39) \cdot 10^{-4}$ & $(1.55 \pm 0.34) \cdot 10^{-4}$ &  258.8 $\pm$ 11.0 &   Glau. MC   & 256.0 $\pm$ 3.9   & \cite{Adamczewski18}\\
 Au+Au & 2.415 &  1.23    &  20   -  40   & $(1.28 \pm 0.14) \cdot 10^{-4}$ & $(5.3  \pm 1.0 ) \cdot 10^{-5}$ &  127.1 $\pm$  7.1 &   Glau. MC   & 125.8 $\pm$ 4.2   & \cite{Adamczewski18}\\
  C+C  & 2.518 &  1.5     &   0   - 100   & $(1.5  \pm 0.6 ) \cdot 10^{-5}$ &          {}                     &      6            &    geom.     &   6.27$\pm$ 0.11  & \cite{Foerster07}   \\	
 Ni+Ni & 2.518 &  1.5     &   0   -   4.4 & $(6.2  \pm 1.1 ) \cdot 10^{-4}$ &          {}                     &    101            &   Glau. Opt  &  89.71$\pm$ 0.66  & \cite{Foerster07}   \\
 Ni+Ni & 2.518 &  1.5     &   4.4 -  15   & $(5.43 \pm 0.68) \cdot 10^{-4}$ &          {}                     &     77            &   Glau. Opt  &  70.85$\pm$ 0.62  & \cite{Foerster07}   \\
 Ni+Ni & 2.518 &  1.5     &  15   -  26.5 & $(3.10 \pm 0.49) \cdot 10^{-4}$ &          {}                     &     52.8          &   Glau. Opt  &  50.31$\pm$ 0.54  & \cite{Foerster07}   \\
 Ni+Ni & 2.518 &  1.5     &  26.5 -  45.9 & $(1.69 \pm 0.32) \cdot 10^{-4}$ &          {}                     &    31             &   Glau. Opt  &  30.15$\pm$ 0.45  & \cite{Foerster07}   \\
 Ni+Ni & 2.518 &  1.5     &  45.9 - 100   & $(2.30 \pm 0.69) \cdot 10^{-5}$ &          {}                     &     7             &   Glau. Opt  &   8.18$\pm$ 0.18  & \cite{Foerster07}   \\
 Au+Au & 2.518 &  1.5     &   0   -   5.4 & $(2.18 \pm 0.45) \cdot 10^{-3}$ &          {}                     &   336.1           &   Glau. Opt  & 329.2 $\pm$ 3.6   & \cite{Foerster07}   \\
 Au+Au & 2.518 &  1.5     &   5.4 -  18.1 & $(1.89 \pm 0.28) \cdot 10^{-3}$ &          {}                     &   252             &   Glau. Opt  & 241.2 $\pm$ 3.4   & \cite{Foerster07}   \\
 Au+Au & 2.518 &  1.5     &  18.1 -  31.1 & $(1.11 \pm 0.20) \cdot 10^{-3}$ &          {}                     &   165             &   Glau. Opt  & 154.2 $\pm$ 3.8   & \cite{Foerster07}   \\
 Au+Au & 2.518 &  1.5     &  31.1 -  52.3 & $(4.50 \pm 0.96) \cdot 10^{-4}$ &          {}                     &    88             &   Glau. Opt  &  80.7 $\pm$ 3.6   & \cite{Foerster07}   \\ 
 Ar+KCl& 2.611 &  1.756   &   0   -  35   & $(7.1  \pm 1.5 ) \cdot 10^{-4}$ & $(2.60 \pm 0.73) \cdot 10^{-4}$ &    38.5           &    geom.     &  37.17$\pm$ 0.59  & \cite{Agakishiev09} \\
  C+C  & 2.627 &  1.8     &   0   - 100   & $(8.1  \pm 1.3 ) \cdot 10^{-5}$ &          {}                     &     6             &    geom.     &   6.40$\pm$ 0.11  & \cite{Foerster07}   \\
 Ni+Ni & 2.627 &  1.8     &   0   - 11.4  & $(3.01 \pm 0.68) \cdot 10^{-3}$ &          {}                     &    85.9           &    geom.     &  81.52$\pm$ 0.59  & \cite{Barth97}      \\
 Ni+Ni & 2.627 &  1.8     &  11.4 - 17.9  & $(1.72 \pm 0.40) \cdot 10^{-3}$ &          {}                     &    61.4           &    geom.     &  61.64$\pm$ 0.54  & \cite{Barth97}      \\ 
 Ni+Ni & 2.627 &  1.8     &  17.9 - 41.0  & $(6.84 \pm 1.69) \cdot 10^{-4}$ &          {}                     &    37.3           &    geom.     &  38.77$\pm$ 0.45  & \cite{Barth97}      \\
 Ni+Ni & 2.627 &  1.8     &  41.0 - 59.3  & $(1.39 \pm 0.36) \cdot 10^{-4}$ &          {}                     &    15.5           &    geom.     &  18.11$\pm$ 0.36  & \cite{Barth97}      \\
 Al+Al & 2.666 &  1.91    &   0   -  8.6  & $(1.08 \pm 0.17) \cdot 10^{-3}$ & $(3.1  \pm 0.8) \cdot 10^{-4}$ &    42             &    geom.     &  36.8 $\pm$ 1.1   & \cite{Gasik16}      \\
 Ni+Ni & 2.666 &  1.91    &   0   - 56    & $(9.1  \pm 0.28) \cdot 10^{-4}$ & $(4.4  \pm 1.5 ) \cdot 10^{-4}$ &    46.5           &    geom.     &  44.81$\pm$ 0.54  & \cite{Piasecki19}   \\
 Ni+Ni & 2.673 &  1.93    &   0   - 21.4  & $(2.1  \pm 0.20) \cdot 10^{-3}$ &          {}                     &    72.7           &    geom.     &  70.89$\pm$ 0.69  & \cite{Menzel00}     \\
 Ni+Ni & 2.673 &  1.93    &   0   - 22    &        {}                       & $(8.6  \pm  2.2) \cdot 10^{-4}$ &    74             &    geom.     &  70.29$\pm$ 0.69  & \cite{Piasecki16}   \\
  C+C  & 2.698 &  2.0     &   0   - 100   & $(1.56 \pm 0.33) \cdot 10^{-4}$ &          {}                     &     6             &    geom.     &   6.47$\pm$ 0.10  & \cite{Foerster07}   \\
 Au+Au & 3.000 &  2.92    &   0   -  10   & $(8.70 \pm 0.55) \cdot 10^{-2}$ & $(2.01 \pm 0.52) \cdot 10^{-2}$ &   305.2           &   Glau. MC   & 311.0 $\pm$ 2.3   & \cite{Abdallah22}   \\
 Au+Au & 3.000 &  2.92    &  10   -  40   & $(3.39 \pm 0.21) \cdot 10^{-2}$ & $(8.5  \pm 2.1 ) \cdot 10^{-3}$ &   150.2           &   Glau. MC   & 161.4 $\pm$ 3.5   & \cite{Abdallah22}   \\
 Au+Au & 3.000 &  2.92    &  40   -  60   & $(7.9  \pm 0.7 ) \cdot 10^{-3}$ & $(2.6  \pm 0.7 ) \cdot 10^{-3}$ &    47.9           &   Glau. MC   &  57.8 $\pm$ 3.3   & \cite{Abdallah22}   \\
 Au+Au & 4.859 &  10.7    &   0   -   5   & $ 4.14 \pm 0.09 $               &          {}                     &   354             &    geom.     & 341.5 $\pm$ 3.8   & \cite{Ahle98}       \\
 Au+Au & 4.859 &  10.7    &   5   -  12   & $ 3.29 \pm 0.08 $               &          {}                     &   312             &    geom.     & 278.2 $\pm$ 3.9   & \cite{Ahle98}       \\
 Au+Au & 4.859 &  10.7    &  12   -  23   & $ 2.23 \pm 0.04 $               &          {}                     &   248             &    geom.     & 205.7 $\pm$ 4.2   & \cite{Ahle98}       \\
 Au+Au & 4.859 &  10.7    &  23   -  39   & $ 1.28 \pm 0.03 $               &          {}                     &   164             &    geom.     & 128.1 $\pm$ 4.6   & \cite{Ahle98}       \\
 Au+Au & 4.859 &  10.7    &  39   -  75   & $ 0.34 \pm 0.02 $               &          {}                     &    62             &    geom.     &  44.4 $\pm$ 2.8   & \cite{Ahle98}       \\
  \hline
 \end{tabular}
 \caption{Yields of K$^-$ and $\phi$ mesons emitted from nuclei colliding at given available energies (as well as beam kinetic energies) and centrality ranges. 
          Values of \Apart are shown as published originally and according to our Glauber Monte Carlo analysis (see text for details.)}
 \label{tab:yields_kmphi}
\end{table}

\begin{table}[htb]
 \setlength{\tabcolsep}{4.5pt}
 \centering
 \begin{tabular}{|c|c|c|c|c|c|c|c|c|c|}
  \hline
  {}   & \sqSnn&$\Tbeam/A$&  Centrality  &      \multicolumn{2}{c|}{Yields}          & \multicolumn{2}{c|}{As published}&    This work      &          {}         \\
System & [GeV] &  [GeV]  &     [\%]      &        K$^0_s$     & $\Lambda + \Sigma^0$ &     \Apart        &     Model    &     \Apart        &         Ref.        \\
  \hline
 Au+Au & 2.415 &  1.23   &   0   -  10   & 0.0284 $\pm$ 0.0025 & 0.082  $\pm$ 0.007  &  303   $\pm$ 11.0 &   Glau. MC   &  300.4 $\pm$ 4.0  & \cite{Adamczewski19}\\
 Au+Au & 2.415 &  1.23   &  10   -  20   & 0.0158 $\pm$ 0.0013 & 0.049  $\pm$ 0.005  &  213.1 $\pm$ 11.1 &   Glau. MC   &  212.3 $\pm$ 3.8  & \cite{Adamczewski19}\\
 Au+Au & 2.415 &  1.23   &  20   -  30   & 0.0106 $\pm$ 0.0009 & 0.0317 $\pm$ 0.0026 &  149.8 $\pm$  9.7 &   Glau. MC   &  149.3 $\pm$ 4.2  & \cite{Adamczewski19}\\
 Au+Au & 2.415 &  1.23   &  30   -  40   & 0.0066 $\pm$ 0.0006 & 0.019  $\pm$ 0.002  &  103.1 $\pm$  6.8 &   Glau. MC   &  102.2 $\pm$ 4.2  & \cite{Adamczewski19}\\
 Ar+KCl& 2.611 &  1.756  &   0   -  35   & 0.0115 $\pm$ 0.001  & 0.0409 $\pm$ 0.0033 &   38.5            &    geom.     &   37.17$\pm$ 0.59 & \cite{Agakishiev10} \\
 Ar+KCl& 2.627 &  1.8    &   0   -  10   &         {}          & 0.042  $\pm$ 0.012  &         {}        &      {}      &   52.74$\pm$ 0.77 & \cite{Harris81}     \\ 
 Al+Al & 2.666 &  1.91   &   0   -  20   & 0.019  $\pm$ 0.002  &          {}         &         {}        &      {}      &   31.32$\pm$ 1.04 & \cite{Lopez10}      \\ 
 Ni+Ni & 2.673 &  1.93   &   0   -   3.6 & 0.052  $\pm$ 0.005  & 0.153  $\pm$ 0.014  &   94              &    geom.     &   93.01$\pm$ 0.66 & \cite{Merschmeyer07}\\
 Ni+Ni & 2.673 &  1.93   &   3.6 -  14.3 & 0.034  $\pm$ 0.004  & 0.112  $\pm$ 0.012  &   63              &    geom.     &   74.38$\pm$ 0.84 & \cite{Merschmeyer07}\\
 Ni+Ni & 2.673 &  1.93   &  14.3 -  25.0 & 0.021  $\pm$ 0.005  & 0.054  $\pm$ 0.008  &   43              &    geom.     &   53.19$\pm$ 0.82 & \cite{Merschmeyer07}\\
 Ni+Ni & 2.673 &  1.93   &  25.0 -  37.1 & 0.012  $\pm$ 0.006  & 0.033  $\pm$ 0.008  &   32              &    geom.     &   36.70$\pm$ 0.69 & \cite{Merschmeyer07}\\
 Ni+Cu & 2.687 &  1.97   &   0   - 100   &         {}          & 0.037  $\pm$ 0.008  &         {}        &      {}      &   28.17$\pm$ 0.35 & \cite{Justice98}    \\
 Au+Au & 2.698 &  2.0    &   0   -   5   & 0.18   $\pm$ 0.02   & 0.58   $\pm$ 0.04   &         {}        &      {}      &  335.6 $\pm$ 3.5  & \cite{Pinkenburg02} \\
 Au+Au & 4.859 &  10.7   &   0   -   5   &         {}          & 16.7   $\pm$ 0.5    &   354             &    geom.     &  341.5 $\pm$ 3.8  & \cite{Ahle98}       \\
  \hline
 \end{tabular}
 \caption{Yields of K$^0$ and $\Lambda$ emitted from nuclei colliding at given available energies (as well as beam kinetic energies) and centrality ranges. 
          Values of \Apart are shown as published originally and according to our Glauber Monte Carlo analysis (see text for details.)}
 \label{tab:yields_k0Lam}
\end{table}

The procedure performed for each colliding system at specific centrality range 
starts with a generation of $4 \cdot 10^5$ collision events,
where for each event a set of positions of nucleons within nuclei 
is randomly generated according to an assumed density profile.
A nucleon from a nucleus A is counted as {\it participant} if it overlaps 
with at least one nucleon from a nucleus B. The {\it overlap} is defined 
such that the centers of these nucleons are closer than 
$\sqrt{\sigma_\mathrm{NN,inel} / \pi}$, where $\sigma_\mathrm{NN,inel}$ 
is the cross section for the inelastic NN collision.
This procedure gives a distribution of number of participants ($N_\mathrm{part}$).
For each experimental data point, a given centrality range is mapped 
into the range of $N_\mathrm{part}$. An averaged $N_\mathrm{part}$ 
within this group is identified as \Apart.
Two kinds of input to this procedure are necessary:

\begin{itemize}
 \item value of $\sigma_\mathrm{NN,inel}$ at a given \sqSnn,
 \item nuclear density profile.
\end{itemize}

\noindent For each experimental data point of strangeness yield 
the $\sigma_\mathrm{NN,inel}$ was obtained as an isospin-averaged combination 
of the cross sections of pp (= nn) and pn interaction,

\begin{equation}
 \sigma_\mathrm{NN} ~=~ 
 \frac{(Z_\mathrm{p} Z_\mathrm{t} + N_\mathrm{p} N_\mathrm{t}) 
 \cdot \sigma_\mathrm{pp} ~+~ (Z_\mathrm{p} N_\mathrm{t} + N_\mathrm{p} Z_\mathrm{t}) \cdot \sigma_\mathrm{np(pn)}}
 {A_\mathrm{p} A_\mathrm{t}}
\end{equation}

\noindent where p and t subscripts correspond to projectile and target,
and $Z$, $N$ and $A$ are the number of protons, neutrons and nucleons.
The excitation functions of both the pp and pn (np) channels were the 
result of the $\sigma_\mathrm{total} - \sigma_\mathrm{elastic}$ 
subtraction of the data from \cite{PDG}. Results of such a subtraction
were already shown in Figs. 77 and 78 of \cite{BehruzMsC}. 
However, it has to be noted that the experimentally extracted
values of np cross sections differ a bit from those from pn experiments.
From this database, the values of elementary cross sections were
interpolated at given \sqSnn via the polynomial fit to the neighbouring points.
We found that a precise form of interpolating function and statistical errors 
had a minor impact on the uncertainties. The dominant factors were: 
(1) the experiment-driven differences between $\sigma_\mathrm{pn}$ and
$\sigma_\mathrm{np}$, and (2) a relative scarcity of 
$\sigma_\mathrm{pn}$ data points at $\sqrt{s} \gtrsim 2.25$ GeV,
as well as of $\sigma_\mathrm{np}$ data points at $\sqrt{s} \lesssim 2.3$ GeV. 
These scarcities cause the susceptibility of the extracted values of the 
interpolation to the range of fitting window. To assess this contribution 
to the systematic errors, the highest and lowest variants of the extracted 
cross sections were found for each experiment.

An experimental extraction of the nuclear profile usually depends on the 
assumed functional form. 
The systematics \cite{Vries87,Moller95} and other analyses
\cite{Alver08,Loizides15,Loizides18,Ficenec70,Reuter82,Sick82,Lim19}
provide a broad range of variants of profiles for many nuclei. 
To assess the systematic error contribution to \Apart due to choice 
of a nuclear shape we performed our Glauber MC calculations for the available 
functional forms and - for each form - the dominant uncertainties of values 
of relevant parameters. The details of this procedure are described in 
\mbox{Appendix A}.
The values of \Apart obtained within this procedure are shown in Tables 
\ref{tab:yields_kp}, \ref{tab:yields_kmphi} and \ref{tab:yields_k0Lam}. 
The statistical uncertainties are negligible. 
The systematic ones stem from previously discussed uncertainties of 
$\sigma_\mathrm{NN,inel}$ and of nuclear density profiles. 
As an uncertainty of each point we took half of distance between the 
maximal and minimal result due to variations of these contributions. 

\begin{figure}
  \centering
  \includegraphics[width=\textwidth]{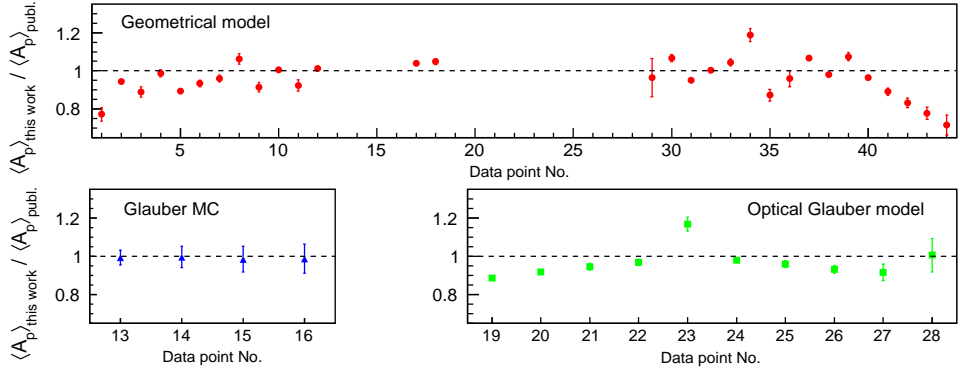}
  \caption{Ratio of \Apart values for K$^+$ data obtained in this work to 
  those published in the original papers. Numbering scheme of data points is 
  as in Table \ref{tab:yields_kp}. Full dots correspond to data where the 
  geometrical model was originally used, squares refer to the Optical Glauber, 
  and triangles - to the Glauber MC. }
  \label{fig:GeoGlauMCK+}
\end{figure}

To shed light on the extent of changes that our method introduces, 
the ratios of corrected to originally published \Apart values for K$^+$ 
data are shown in Fig.~\ref{fig:GeoGlauMCK+}, 
where the numbering of data points is the one from the Table \ref{tab:yields_kp}. 
The subsequent panels on this figure correspond to three different 
models estimating \Apart, used in the original works. 
One observes that corrections are largest for the data obtained 
originally from the geometrical model.
They often exceed 10\% and in some cases 25\%. For the data where 
the Optical Glauber approach was originally used, the deviation is 
usually smaller, but still in one case reaches 20\%. 
For the data where \Apart was originally found using the Glauber MC 
approach, we are in good agreement with the original results, 
which supports our calculations. 
We have also verified that the RMS values of the density profiles 
obtained in our analysis are in agreement with the experimental ones, 
reported in \cite{Landolt04}.

\section{Parametrization of yields of strange hadrons}
\label{Sect:paramcurve}

Our aim is to find the type of curve that fits best to the available
experimental yields of K$^{\pm,0}$, $\phi$ and $\Lambda$ 
at \sqSnn between 2.16 and 3.0 GeV. As we show further, the AGS yields 
at 4.86 GeV deviate already too far from the rest of considered points. 
The curve should serve as a convenient tool to predict the yield at given 
\sqSnn and \mbox{\Apart,} providing also its uncertainty. 
To keep the numerical stability of these results, such a curve should 
have as few parameters, as possible. We selected the function type
which factorizes the mass scaling as the power law and assumes 
the exponential dependence on available energy:

\begin{equation}
  P \left( \Apart ,~ \sqSnn \right) ~=~ 
    N ~\Apart^\alpha ~\exp \left[ - \left(C \cdot \sqSnn \right)^{~\beta} \right]    \quad\quad .
  \label{Eq:fitcurve}
\end{equation}

\noindent We found the $\beta$ parameters negative for every studied hadron type, 
therefore the curve has a saturating behaviour at large $\sqSnn$. 
Despite this feature, the formula appears to fit better to the data than 
the power function, $P \sim \sqSnn^\beta$.
We found that for K$^+$ and K$^-$ the fitting procedure with all the four 
parameters left free generally worked well and the predictions 
of specific yields delivered rather reasonable results with limited uncertainty,
which will be discussed further.
However, the meager statistics in cases of the other hadrons caused
problems with the covariance matrices. Therefore, for K$^0_s$, \mbox{$\Lambda+\Sigma^0$}
and $\phi$ we limited the parameter space by fixing $C$ first,
and adjusting its value manually, following the requirement of best 
$\chi^2/\nu$ of the 3-parameter fit. 
We checked the effect of this procedure for K$^+$ and K$^-$, where both 
approaches were possible. We found that the values of predicted yields are
identical up to third significant digit, but the fit in 3-parameter space
generated uncertainties about 3 times lower than the full fit.
It suggests that uncertainties for yields of K$^0_s$, $\Lambda+\Sigma^0$
and $\phi$ are probably also underestimated. 
We admit that it is a temporary solution in light of the limited statistics,
and welcome any new data points to elevate the procedure to the full approach.

The parameters of the best fits of Eq.~\ref{Eq:fitcurve} to the experimental
yields of considered hadrons are shown in Table~\ref{tab:fitresults}. 
The full covariance matrices are presented in \mbox{Appendix B}.
The $\chi^2/\nu$ values appear to be reasonable. For the case of K$^+$, 
where $\chi^2/\nu$ reaches a relatively high value of 3.6,
we note that the removal of point No. 34 in Table~\ref{tab:yields_kp} 
results in a considerable drop of $\chi^2/\nu$ down to 2.1. 
However, we cannot pinpoint an experimental defect to substantiate 
the removal of this point. 
Interestingly, the $\alpha$ exponent of the yield dependence on \Apart appears 
to be identical within uncertainties for K$^+$ and K$^-$. This result confirms 
and sharpens the finding of a constant K$^-$/K$^+$ ratio as a function of 
\Apart, obtained for the smaller dataset of Au+Au at \sqSnn = 2.52 GeV 
and Ni+Ni at 2.52 and 2.67 GeV~\cite{Foerster07}.
A relative goodness of fitting the Eq.~\ref{Eq:fitcurve} to the data 
also suggests, that the constancy of $\alpha$ exponent across 
the studied range of \sqSnn may be a good hypothesis. 
We shall examine this issue further in the paper.
A strength of the $\beta$ parameter steers the steepness of the curve
with $\sqSnn$. We note that for $\Lambda+\Sigma^0$ the best fit value
corresponds to an unusually high rise, and causes that for the experimental 
points at highest energy the curve is nearly saturated. Clearly more data 
is needed to verify this parametrization, and in particular the extrapolations 
of this curve seem unreliable, as it will be shown further. 

For K$^+$ and K$^-$ the $N$, $\beta$ and $C$ parameters exhibit strong 
(anti-)correlations (c.f. Tab.~\ref{tab:covmats}). 
As a result, despite the fact that $N$ parameters are burdened with large
uncertainties, the non-diagonal terms of the covariance matrix
strongly reduce the uncertainties of the estimated yields. 

\begin{figure*}
 \begin{minipage}[c]{0.495\textwidth}
  \begin{center}
   \includegraphics[width=0.97\textwidth]{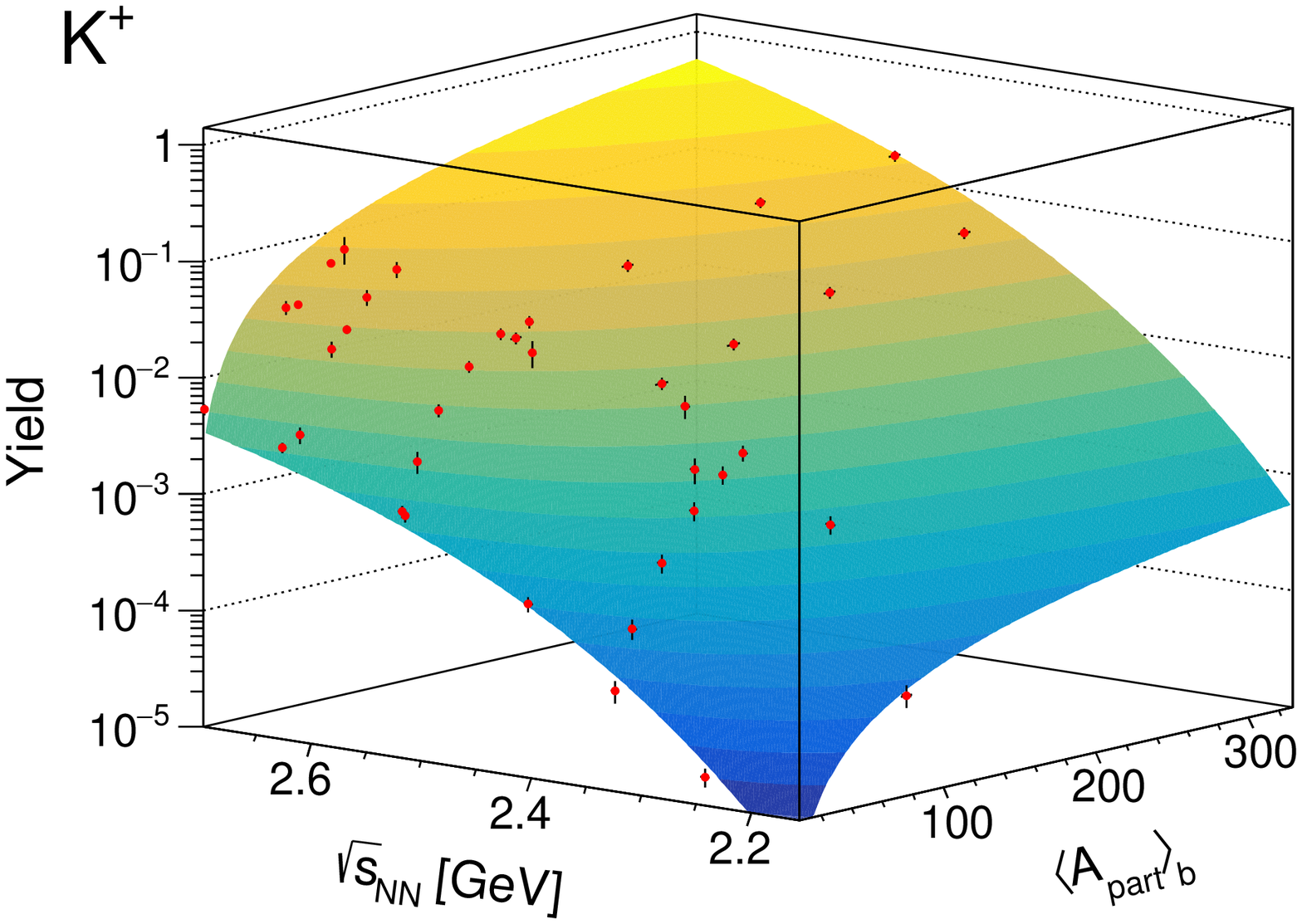}
  \end{center}
 \end{minipage}
 \begin{minipage}[c]{0.495\textwidth}
  \begin{center}
   \includegraphics[width=0.97\textwidth]{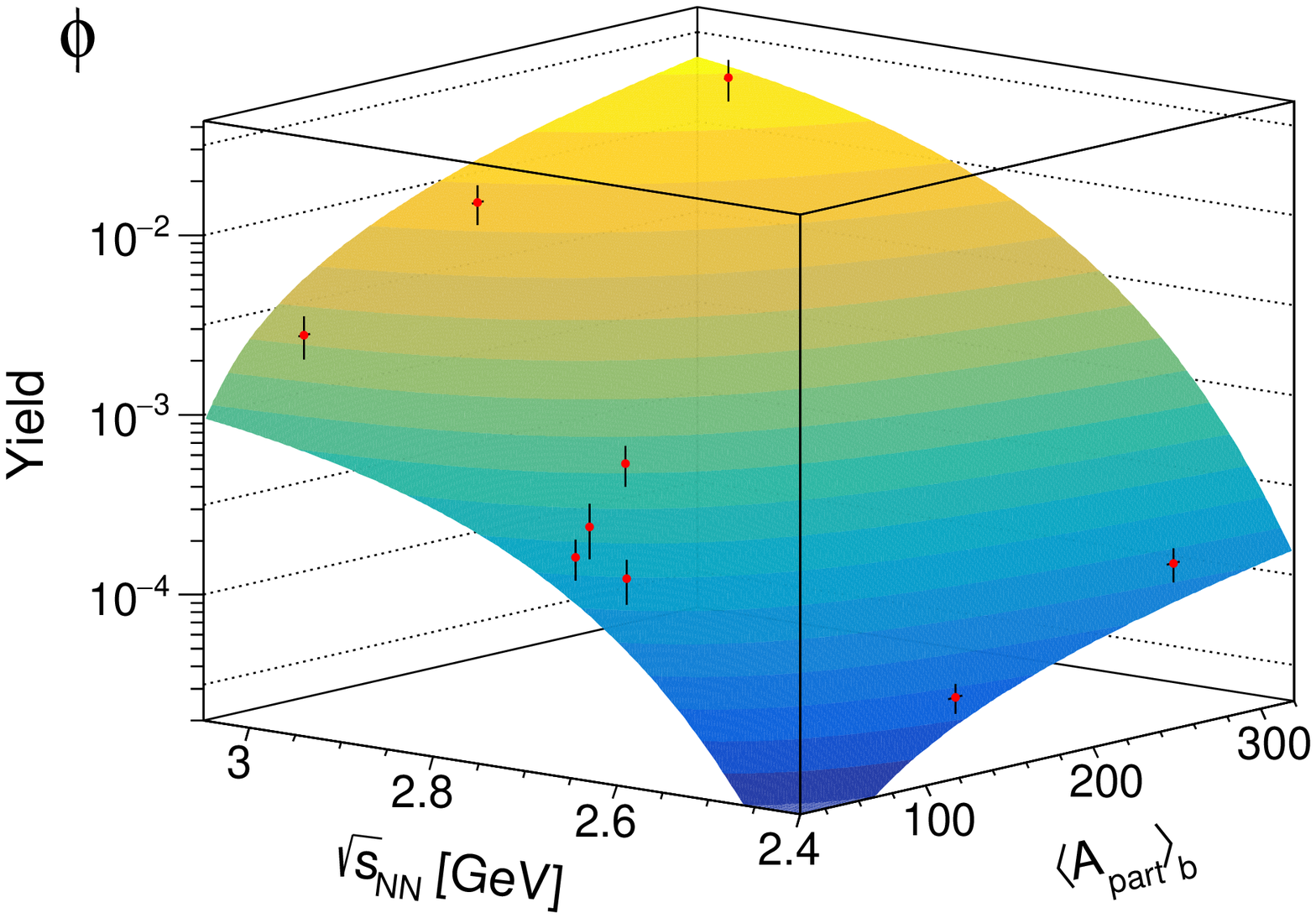}
  \end{center}
 \end{minipage}
 \begin{minipage}[c]{0.495\textwidth}
  \begin{center}
   \includegraphics[width=0.97\textwidth]{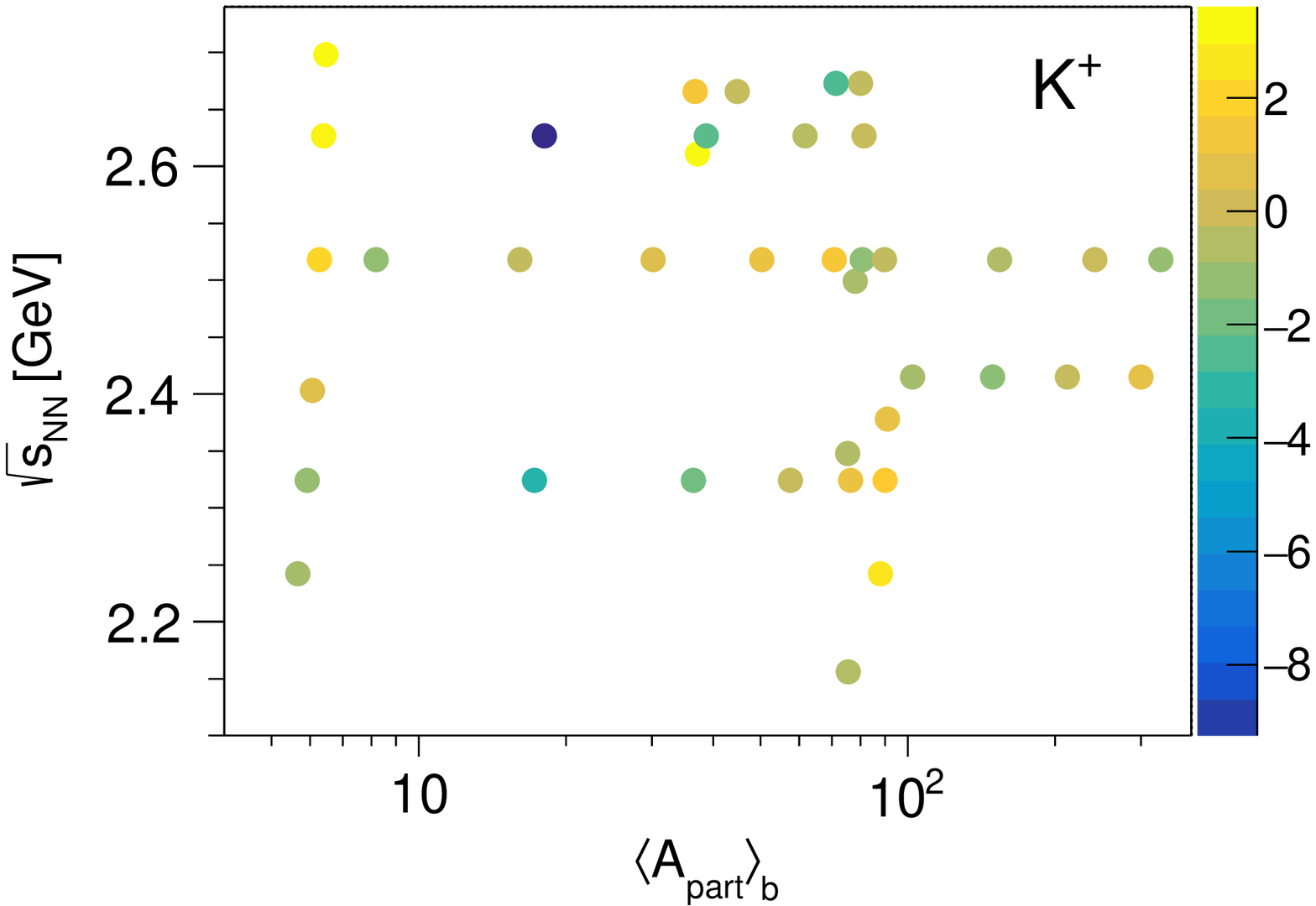}
  \end{center}
 \end{minipage}
 \begin{minipage}[c]{0.495\textwidth}
  \begin{center}
   \includegraphics[width=0.97\textwidth]{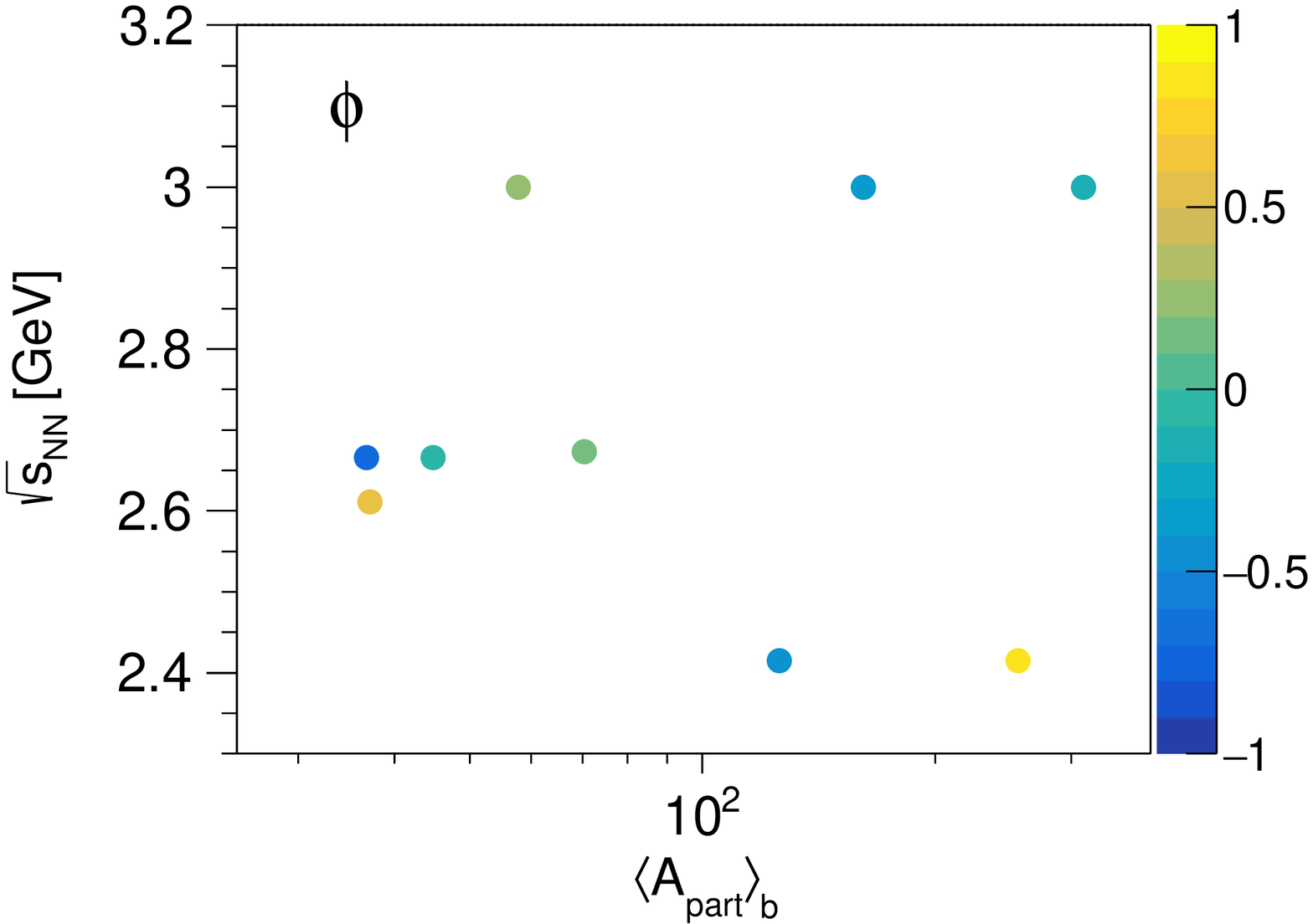}
  \end{center}
 \end{minipage}
 \caption{\label{fig:param3D}(Top panels) systematics of yields of K$^+$ and 
 $\phi$ mesons emitted from heavy-ion collisions in the near-threshold range, 
 as function of available energy in the NN channel, \sqSnn, and mean number 
 of participants, \Apart.
 The points depict the experimental data. The curves are the best fits of 
 Eq.~\ref{Eq:fitcurve} to these data. (Bottom panels:) the maps of standard
 deviations of the data points from the fitted curves.}
\end{figure*}

Fig.~\ref{fig:param3D} visualizes the experimental data points and the best fit 
curves describing the yields as a function of \sqSnn and \Apart for two cases
of, respectively, the richest statistics (K$^+$) and the poorest one ($\phi$). 
A rather good factorization into these two quantities can be observed.
For orientation, the standard deviations of experimental points from their
fit predictions are shown in the lower panels of this figure.

\begin{table}[htb]
 \centering
 \begin{tabular}{|c|c|c|c|c|c|}
  \hline
    {}      &          K$^+$               &             K$^-$           &           K$^0_s$             &     $\Lambda + \Sigma^0$    &            $\phi$           \\
  \hline
    $N$     &$(3.0 \pm 0.9) \cdot 10^{-3}$ &$(1.6 \pm 0.7) \cdot 10^{-4}$& $(4.5 \pm 0.9) \cdot 10^{-3}$ &$(5.1 \pm 1.0) \cdot 10^{-4}$&$(2.6 \pm 1.4) \cdot 10^{-5}$\\
  $\alpha$  &        1.316 $\pm$ 0.021     &     1.316 $\pm$ 0.043       &       1.05 $\pm$ 0.05         &       1.22 $\pm$ 0.04       &        1.27 $\pm$ 0.12      \\
  $\beta$   &        -6.18 $\pm$ 0.5       &    -7.2   $\pm$ 0.7         &     -5.71  $\pm$ 0.10         &     -67    $\pm$ 6          &       -9.97 $\pm$ 0.23      \\
    $C$     &        0.323 $\pm$ 0.010     &     0.320 $\pm$ 0.009       &        0.320 (fixed)          &         0.41 (fixed)        &         0.353 (fixed)       \\
        \hline
$\chi^2/\nu$&             3.6              &              2.2            &              2.0              &             1.4             &             0.15            \\
        \hline
 \end{tabular}
 \caption{Results of fit of curve \ref{Eq:fitcurve} to the yields of strange hadrons 
   as function of \Apart and \sqSnn. For K$^\pm$ all the parameters were fitted.
   For the other hadrons $C$ was fixed during the fit. The data points are listed in 
   Tables \ref{tab:yields_kp}, \ref{tab:yields_kmphi} and \ref{tab:yields_k0Lam}. 
   For prediction of yields see the full covariance matrices in Tab.~\ref{tab:covmats} 
   and discussion in text.}
 \label{tab:fitresults}
\end{table}

We shall now move to the predictions of yields with the found curves. 
The estimations of uncertainties for a given hadron type can be 
usually done via the error propagation formula including the covariance terms.
We cross-checked this method by the bootstrap-like random pulling 
of curve parameters from the multivariate Gaussian distribution 
parametrized by the covariance matrix obtained from the fit. 
We found that the probability distributions of some of yields had 
non-Gaussian, asymmetric tails. Therefore, we modified the definition of
the yield uncertainty to be: half of distance keeping central 68.3\% 
of the probability distribution generated in such a way. 
In some cases we found the uncertainties up to 2 times larger than those 
calculated with the error propagation formula. All the uncertainties 
reported in this work were obtained using the bootstrap method.

Our first aim is to compare with the experimental data the interpolations 
at the lowest available energies (2.16 GeV for K$^+$ and 2.41 GeV for the 
other hadrons) and at the highest one (3 GeV) used for fitting. 
As it is seen in the upper panels of Table~\ref{tab:fitpredictions},
the agreement is robust.
We also compare the interpolations to the experimental data for 
35\% most central Ar+KCl collisions at \sqSnn = 2.61 GeV, as this is the only 
experiment in the studied energy range, where the yields of all five strange 
hadrons are available at the same centrality range. Here, for the most of 
yields a very good agreement is found, with the exception of K$^+$ 
where the distance is about 3.3 standard deviations. 
We also note that the relative errors of the predictions themselves remain 
within 4-20\% range.

\begin{table}[tb]
 \setlength{\tabcolsep}{4pt}
 \centering
 \begin{tabular}{|c|c|c|c|c|c|c|c|c|c|}
 \hline
\sqSnn &$\Tbeam/A$&\Apart& System and   &Type&       K$^+$         &      K$^-$         &       K$^0_s$   &$\Lambda+\Sigma^0$&      $\phi$      \\
{[GeV]}& {[GeV]}  &  {}  &centrality [\%]&{} &        {}           &        {}          &        {}       &        {}        &       {}         \\
 \hline
   {}  &    {}    &  {}  &   Au + Au    & {} &   $\times 10^{-3}$  &  $\times 10^{-4}$  & $\times 10^{-2}$& $\times 10^{-2}$& $\times 10^{-5}$  \\
  2.16 &   0.6    & 75.7 &   0 - 100    & fit& 0.077  $\pm$ 0.016  &        {}          &        {}       &        {}       &        {}         \\
   {}  &    {}    &  {}  &      {}      & exp& 0.073  $\pm$ 0.016  &        {}          &        {}       &        {}       &        {}         \\
  2.41 &  1.23    &149.3 &   20 - 30    & fit& 20.9   $\pm$ 0.9    &        {}          & 1.10 $\pm$ 0.06 & 3.22 $\pm$ 0.38 &        {}         \\
   {}  &   {}     &  {}  &      {}      & exp& 18.8   $\pm$ 2.1    &        {}          & 1.06 $\pm$ 0.09 & 3.17 $\pm$ 0.26 &        {}         \\
  2.41 &  1.23    &125.8 &   20 - 40    & fit&         {}          &  1.43  $\pm$ 0.19  &        {}       &        {}       &   5.7  $\pm$ 1.1  \\
   {}  &   {}     &  {}  &      {}      & exp&         {}          &  1.28  $\pm$ 0.14  &        {}       &        {}       &   5.3  $\pm$ 1.0  \\
 \hline
   {}  &    {}    &  {}  &   Ar + KCl   & {} &   $\times 10^{-2}$  &  $\times 10^{-4}$  & $\times 10^{-2}$& $\times 10^{-2}$& $\times 10^{-4}$  \\
  2.61 &  1.756   & 37.2 &   0 - 35     & fit& 1.98   $\pm$ 0.07   &  4.78  $\pm$ 0.33  & 1.23 $\pm$ 0.07 & 4.11 $\pm$ 0.23 &  2.21  $\pm$ 0.39 \\
   {}  &   {}     &  {}  &      {}      & exp& 2.80   $\pm$ 0.24   &  7.1   $\pm$ 1.5   & 1.15 $\pm$ 0.10 & 4.09 $\pm$ 0.33 &  2.6   $\pm$ 0.7  \\
 \hline
   {}  &   {}      &  {}  &   Au + Au    &  {}&      $\times 1$     &  $\times 10^{-1}$  &        {}       &   $\times 1$    & $\times 10^{-3}$ \\
  3.0  &  2.92     &161.4 &   10 - 40    & fit&         {}          &  0.341 $\pm$ 0.033 &        {}       &        {}       &   8.9 $\pm$ 1.7  \\
   {}  &   {}      &  {}  &      {}      & exp&         {}          &  0.339 $\pm$ 0.021 &        {}       &        {}       &   8.5 $\pm$ 2.1  \\
  4.86 &  10.7     &341.5 &    0 - 5     & fit&   6.0  $\pm$ 1.8    &   3.4  $\pm$ 0.8   &        {}       & 0.617$\pm$ 0.039&       {}         \\
   {}  &   {}      &  {}  &      {}      & exp& 24.2   $\pm$ 0.9    &  41.4  $\pm$ 0.9   &        {}       & 16.7 $\pm$ 0.5  &       {}         \\
 \hline
 \hline
   {}  &   {}      &  {}  &   Au + Au    & fit&   $\times 10^{-4}$  &  $\times 10^{-7}$  & $\times 10^{-4}$& $\times 10^{-9}$& $\times 10^{-9}$ \\
  2.16 &  0.6      &251.7 &    0 - 10    &    & 3.7   $\pm$ 0.8     &  1     $\pm$ 1     & 3.6  $\pm$ 0.8  &      $< 1$      &   2   $\pm$ 2    \\
   {}  &   {}      &174.8 &   10 - 20    &    & 2.3   $\pm$ 0.5     &  0.6   $\pm$ 0.7   & 2.5  $\pm$ 0.6  &      $< 1$      &   1   $\pm$ 1    \\
 \hline
   {}  &   {}      &  {}  &   Au + Au    & fit&   $\times 10^{-3}$  &  $\times 10^{-6}$  & $\times 10^{-3}$&$\times 10^{-10}$& $\times 10^{-7}$ \\
  2.24 &  0.8      &290.0 &    0 - 10    &    & 3.32  $\pm$ 0.37    &  4.1   $\pm$ 2.4   & 2.21 $\pm$ 0.32 &      $< 1$       &   4.5 $\pm$ 2.4 \\
   {}  &   {}      &203.6 &   10 - 20    &    & 2.08  $\pm$ 0.23    &  2.6   $\pm$ 1.5   & 1.52 $\pm$ 0.23 &      $< 1$       &   2.9 $\pm$ 1.5 \\
 \hline
   {}  &   {}      &  {}  &   Ag + Ag    & fit&   $\times 10^{-2}$  &  $\times 10^{-4}$  & $\times 10^{-2}$& $\times 10^{-2}$& $\times 10^{-5}$ \\
  2.41 &  1.23     &156.0 &    0 - 10    &    & 2.21  $\pm$ 0.09    &  1.90  $\pm$ 0.24  & 1.16 $\pm$ 0.07 & 3.40 $\pm$ 0.40 &   7.5 $\pm$ 1.5  \\
   {}  &   {}      &111.3 &   10 - 20    &    & 1.42  $\pm$ 0.06    &  1.22  $\pm$ 0.16  & 0.81 $\pm$ 0.05 & 2.25 $\pm$ 0.27 &   4.9 $\pm$ 1.0  \\
   {}  &   {}      & 79.1 &   20 - 30    &    & 0.904 $\pm$ 0.035   &  0.78  $\pm$ 0.11  & 0.566$\pm$ 0.036& 1.49 $\pm$ 0.18 &   3.2 $\pm$ 0.7  \\
   {}  &   {}      & 55.1 &   30 - 40    &    & 0.562 $\pm$ 0.022   &  0.48  $\pm$ 0.07  & 0.387$\pm$ 0.028& 0.96 $\pm$ 0.12 &   2.0 $\pm$ 0.4  \\
 \hline
  2.55 &  1.58     &159.5 &    0 - 10    & fit&  8.38 $\pm$ 0.34    & 15.7   $\pm$ 1.2   & 3.72 $\pm$ 0.17 & 23.1 $\pm$ 1.3  &   70  $\pm$ 11   \\
   {}  &   {}      &114.1 &   10 - 20    &    &  5.39 $\pm$ 0.20    & 10.1   $\pm$ 0.7   & 2.61 $\pm$ 0.11 & 15.4 $\pm$ 0.8  &   46  $\pm$ 7    \\
   {}  &   {}      & 81.0 &   20 - 30    &    &  3.43 $\pm$ 0.12    &  6.44  $\pm$ 0.45  & 1.82 $\pm$ 0.08 & 10.1 $\pm$ 0.5  &   30  $\pm$ 5    \\
   {}  &   {}      & 56.2 &   30 - 40    &    &  2.12 $\pm$ 0.07    &  3.99  $\pm$ 0.29  & 1.24 $\pm$ 0.06 & 6.50 $\pm$ 0.34 &   18.7$\pm$ 3.0  \\
 \hline
   {}  &   {}      &  {}  &   Au + Au    & fit&      $\times 1$     &  $\times 10^{-2}$  & $\times 10^{-1}$& $\times 10^{-1}$& $\times 10^{-2}$ \\
  3.0  &  2.92     &311.0 &    0 - 10    &    & 1.68  $\pm$ 0.20    &  8.1   $\pm$ 0.8   & 5.3  $\pm$ 0.6  & 5.50 $\pm$ 0.33 &  2.0  $\pm$ 0.4  \\
   {}  &   {}      &161.4 &   10 - 40    &    & 0.71  $\pm$ 0.08    &  3.41  $\pm$ 0.33  & 2.6  $\pm$ 0.23 & 2.48 $\pm$ 0.12 &  0.89 $\pm$ 0.17 \\
   {}  &   {}      & 57.8 &   40 - 60    &    & 0.184 $\pm$ 0.019   &  0.88  $\pm$ 0.09  & 0.90 $\pm$ 0.07 & 0.710$\pm$ 0.033&  0.24 $\pm$ 0.04 \\
 \hline
   {}  &   {}      &100.0 &   Au + Au    & fit&   $\times 10^{-1}$  &  $\times 10^{-2}$  & $\times 10^{-1}$& $\times 10^{-1}$& $\times 10^{-3}$ \\
  2.70 &  2.0      &  {}  &      {}      &    &  1.23  $\pm$ 0.06   &  0.381$\pm$ 0.29   & 0.558$\pm$ 0.032& 1.38 $\pm$ 0.06 & 1.52 $\pm$ 0.24  \\
  3.32 &  4.0      &  {}  &      {}      &    &  6.7   $\pm$ 1.1    &  3.7  $\pm$ 0.5    & 2.78 $\pm$ 0.24 & 1.38 $\pm$ 0.06 & 7.2  $\pm$ 1.2   \\
  3.85 &  6.0      &  {}  &      {}      &    &  9.8   $\pm$ 2.3    &  5.6  $\pm$ 1.5    & 4.15 $\pm$ 0.38 & 1.38 $\pm$ 0.06 & 8.5  $\pm$ 1.4   \\
 \hline
 \end{tabular}
 \caption{Experimental and predicted yields of strange hadrons 
 for selected beam energies and centralities.  
 See text for the choice of these points and discussion.}
 \label{tab:fitpredictions}
\end{table}

\subsection{Predictions for the upcoming data and planned experiments}

The overall positive results of comparisons of the parametrizations
to the experimental yields encourages us to present several predictions
for the colliding systems for which the experimental data for strangeness
are currently analysed or planned to be obtained. 
All the \Apart values were calculated with the method described 
in Sect.~\ref{Sect:Apart}. 

Let us first look at the collisions of Ag+Ag at the available energies of 
2.41 and 2.55 GeV (current analysis by HADES). Here, the uncertainties of 
the predicted yields range from about 3\% for K$^+$ down to 22\% 
for $\phi$. The predictions appear to be in good agreement with 
the experimental data in the neighbouring region of \sqSnn and \Apart.

Focusing on collisions of Au+Au at \sqSnn = 2.16 and 2.24 GeV,
being the topic of the HADES S520 experiment accepted by the GSI,
we notice that these energies probe the edge of available 
data for K$^+$, but fall below the edges for the other hadron types.
Nevertheless one can consider the following ratios of yields: 
K$^0_s$/K$^+$, K$^-$/K$^+$, $\phi$/K$^-$ and $(\Lambda+\Sigma^0)/K^0_s$.

The first ratio reaches on average 1.03$\pm$0.30 at lower energy and 
0.73$\pm$0.17 at the higher one. 
For the closest beam energy where the experimental ratio can be constructed,
\sqSnn of 2.41, the value is within 0.47-0.55 depending on centrality, 
with uncertainty of about 0.07. 
Therefore, our predictions are 1.7 (0.7) standard deviations above the 
compared experimental values for the lower (higher) available energy.
Since the agreement for the K$^+$ yield at 2.16 GeV is very good, 
we conclude that the predictions of K$^0_s$ are reasonable. 
Moving to K$^-$/K$^+$, we first note that for the lower energy the uncertainties
are larger than the yields. For the higher energy, the values of this ratio
are about $(1.2 \pm 0.7) \cdot 10^{-3}$, where the considerable error 
is the consequence of the uncertainty of the K$^-$ yield. 
The experimental value closest to the discussed case is
$7 \cdot 10^{-3}$ for \sqSnn of 2.41 GeV~\cite{Adamczewski18} and
the K$^-$/K$^+$ ratio is known to systematically drop with decreasing the 
available energy (c.f. Fig.~6.2 in~\cite{Hartnack12}).
The ratio predicted at \sqSnn = 2.24 GeV is generally in line with 
the extrapolation of the trend on this plot. 
For the $\phi/$K$^-$ ratio, we notice that at lower energy the uncertainty
of the prediction of $\phi$ yield reaches 100\%. For the higher energy
it is not so drastic, but the ratio of yields, found to be 0.11, is burdened
with 80\% uncertainty. It is not in disagreement with the closest experimental
value of this ratio, $0.46 \pm 0.12$ for \sqSnn = 2.42 GeV~\cite{Adamczewski18},
however, the large uncertainty of our prediction makes it of little use.
The predicted values of the $(\Lambda+\Sigma^0)$ yield are below 10$^{-10}$, 
which is more than 6 orders of magnitude below the yield of K$^0$ and K$^+$.
As the basic expected strangeness production channel is 
\mbox{BB $\rightarrow$ BK$^{+/0}$Y}, where B $\in$ \{N, $\Delta$\}, 
the hyperon and kaon yields should be similar.
This is the case e.g. for \sqSnn = 2.41 GeV, where the ratio is at the 
level \mbox{of 3}. Therefore the predictions for the hyperon production 
at the S520 energies are unreliable,
and probably are the effect of the very steep $\beta$ parameter
found by the fitting procedure, already discussed in the previous section.
To sum up, our predictions at 2.24 GeV seem reasonable for all the kaons,
but at 2.16 GeV - only for K$^+$ and K$^0_s$, while the rest of predictions
is unreliable.

Let us discuss the predictions for \sqSnn = 3 GeV (STAR's lowest 
beam energy) for the currently unpublished yields of K$^+$, K$^0_s$ 
and $\Lambda$ for three centrality classes applied in~\cite{Abdallah22}.
For all the centralities we found K$^-$/K$^+$ ratio to be $(4.8 \pm 0.7) \%$.
This value is well in line with the AGS results obtained at midrapidity 
(c.f. Fig. 3 in \cite{Ahle00}). Since the result from our parametric curve 
for K$^-$ follows the experiment nearly exactly, we conclude that 
the K$^+$ yield is predicted also well. 
The predicted $K^0_s$/K$^+$ ratio rises from 0.32(5) to 0.49(7) while
passing from central to more peripheral collisions. The nearest available energy
and centrality at which such a ratio can be constructed from the 
experimental data is \sqSnn = 2.67 GeV and $\Apart \sim 72$. 
The ratio is within 0.41 and 0.51, depending on the centrality class of 
two adjacent data points for K$^+$. Apparently, the predicted values fit nicely
to the experimental ones at the adjacent energy, hence the prediction for K$^0_s$
is reasonable too.
Let us now focus on the $(\Lambda + \Sigma^0)$/K$^0_s$ ratio.
For the considered centrality classes our prediction points to values 
between 1.13(16) and 0.94(10). However, at \sqSnn of 2.67 GeV
the experimental value of this ratio is found to be between 2.6 and 3.3. 
This discrepancy is probably the consequence of the previously discussed 
reaching the saturation region by Eq.~\ref{Eq:fitcurve} for the $\Lambda$ case.
Hence, we conclude that this prediction is not reliable.
To sum up, among the unpublished yields of K$^+$, K$^0_s$ and 
$\Lambda+\Sigma^0$ at \sqSnn of 3.0 GeV,
the predictions for two former hadrons make sense, but not for the latter one.

The CBM setup, currently being constructed at FAIR, GSI is planned to 
measure heavy-ion collisions at available energies within 2.7 - 5 
GeV~\cite{Cbm17}. Here, we should check, how far in terms of 
available energy our parametrization can be useful.
The experimental yields of K$^\pm$ in this energy region are found 
to be rising sharply, however they were measured only at midrapidity, 
so this data cannot be applied directly to our analysis~\cite{Ahle00}. 
The 4$\pi$ yields of K$^\pm$ and $\Lambda$ are available at 
\sqSnn of 4.86 GeV. 
However, as seen in Table~\ref{tab:fitpredictions}, our parametrization 
strongly underestimates the experimental yields.
The region of available energies on which our fits base, 2.16 - 3.0 GeV, 
is understood to be of (nearly) purely hadronic character. 
At higher available energies the string degrees of freedom 
open up and the chiral symmetry restoration during the 
hadronization process is expected to additionally raise the yields 
(c.f. Sect.~5.2 in \cite{Blei22}).
On the side of fitting, the curve defined by Eq.~\ref{Eq:fitcurve} 
inevitably saturates with energy.
Therefore, an extrapolation of our fit should not go far. 
In Table~\ref{tab:fitpredictions} we present the predictions up to 
\sqSnn = 3.85 GeV only. They were obtained at \Apart = 100.

To assess credibility of these results, let us first examine 
the K$^-$/K$^+$ yield ratio. 
Our fits predict that its values should be 0.031(3), 0.055(12) and 0.057(18), 
respectively from the lowest energy upwards. 
These results can be compared to the experimental excitation function
of this ratio, shown in Fig. 3 of~\cite{Ahle00}, although measured at 
midrapidity. Apparently, two first predictions are in line with
the experimental data, but the third one falls below. 
Regarding the $\phi$/K$^-$ ratio, our curves give the values:
0.40(7), 0.18(4) and 0.15(4) from the lowest energy upwards. 
They seem to nicely follow the experimental trend of this ratio, 
as shown in Fig. 20 of \cite{Song22}. 
Passing to the K$^0_s$/K$^+$ yield ratio, our parametrization proposes,
0.45(3), 0.41(8) and 0.42(11), respectively from the lowest energy upwards.
As mentioned before, the nearest available energy where the experimental 
data is available is 2.67 GeV, and the value of the ratio is located 
between 0.41 and 0.51. All the predictions fit very well to these values. 
Regarding $\Lambda + \Sigma^0$, the predicted yield of these hyperons 
does not change with the beam energy.
It is a consequence of reaching the saturation by this curve, as discussed before.
However, the (\sqSnn, \Apart) coordinates for the lowest CBM energy are situated 
close to two points where the experimental values of $(\Lambda + \Sigma^0)$/K$^0_s$ 
ratio are available. Namely, at (2.67 GeV, 93) and at (2.7 GeV, 336) 
the ratio is found to be 2.9(4) and 3.2(4), respectively. 
Our prediction of the ratio at (2.7 GeV, 100) gives 2.47(18).
As the standard deviations between our prediction and these experimental points
are, respectively, 0.8 and 1.7, we conclude that at the lowest CBM available 
energy of 2.7 GeV the predicted ratio (and consequently the yield of 
$\Lambda + \Sigma^0$) is still reliable. 
To sum up, for the lowest energy all the predictions seem reasonable,
for the middle one - all except $\Lambda + \Sigma^0$ and for the highest one
(at \sqSnn = 3.85 GeV) - only $\phi$/K$^-$ and K$^0_s$/K$^+$ ratios make sense.

Overall, the predictions the K$^+$ and K$^0$ yields seem to be reliable
within the available energy range of [2.16 - 3.0] GeV, whereas for K$^-$
the lower limit should be lifted to 2.24 GeV. For $\Lambda+\Sigma^0$ and $\phi$
the predictions appear to make sense from 2.41 GeV, up to 2.7 (3.0) GeV,
respectively. In terms of the ratios of yields, the predictions for
the K$^-$/K$^+$ seem to be reliable within [2.24 - 3.32] GeV, 
K$^0_s$/K$^+$ makes sense for all the examined range of [2.16 - 3.85] GeV,
and $\phi$/K$^-$ fits the experimental data for [2.41 - 3.85] GeV.

\subsection{Discussion on the $\alpha$ exponent}

A consequence of adopting the parametrization of Eq.~\ref{Eq:fitcurve} 
is the assumption that for every hadron type the $\alpha$ exponent of the 
yield dependency on \Apart does not change with beam energy. 
We tested this assumption by fitting $\alpha$ for groups of data points 
at fixed available energies (or a narrow range of them), 
in the region between 2.3 and 3.0 GeV.
We noticed that the fit of the set of yields of K$^+$ emitted at 
\sqSnn = 2.63 GeV resulted in an unacceptably high $\chi^2/\nu$ of 15. 
Seemingly, the data point for C+C colliding at this energy could not 
be aligned with the yields for Ni+Ni. 
Therefore, we disregarded this dataset.
We also did not consider the $\phi$ meson data, as for the only series 
of points at the same available energy of 2.67 GeV, the result
$\alpha = 1.0 \pm 0.6$ is burdened with too high uncertainty 
to contribute noticeably to the analysis~\cite{Piasecki16}. 
Out of 11 accepted series of data points, 8 were characterized by the same 
available energy within $\pm$~0.01 GeV. For K$^-$, K$^0_s$ and $\Lambda$
emitted from collisions at $\sqSnn \in [2.6 - 2.7]$ GeV we noticed 
some variations of the fitted $\alpha$ values depending on the choice 
of subsample. 
Therefore, for each of these cases we grouped together the data for the
above-mentioned energy range and fitted the Eq.~\ref{Eq:fitcurve} where
the available energy dependence was subject to fitting. 
We then estimated the additional contributions to the systematic errors 
by testing the susceptibility of resulting value of $\alpha$ to the removal 
of any single point from these samples. 

\begin{figure}
  \centering
  \includegraphics[width=0.8\textwidth]{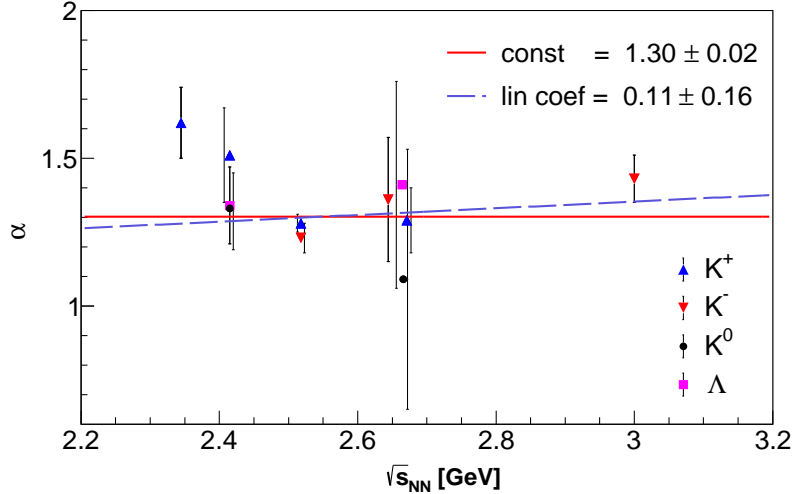}
  \caption{Values of $\alpha$ exponent of Eq.~\ref{Eq:fitcurve}
  obtained for fixed beam energies and given hadron types. 
  To increase readability at clusters of points, some uncertainties
  were shifted to the edges of markers. All the errors are symmetric.
  Dashed (full) lines mark the linear (constant) functions 
  of the best fit to the data.}
  \label{fig:alpha-sqS}
\end{figure}

Figure~\ref{fig:alpha-sqS} shows the results as function of \sqSnn and 
hadron type. Per hadron type we do not observe any statistically meaningful 
linear change of $\alpha$ with \mbox{\sqSnn.} A fit of the linear function
to all the points taken together provides the gradient value of 
$0.11 \pm 0.16$, so the global conclusion is the same.
Therefore, we are encouraged to fit a constant to the whole dataset, 
which gives $\alpha = 1.30 \pm 0.02$. 

We can now compare this result to the $\alpha$ values obtained per hadron, 
shown in Table \ref{tab:fitresults}. Apparently, for K$^+$, K$^-$ and $\phi$ 
mesons the agreement is within 1 standard deviation ($\sigma$), and for 
$\Lambda+\Sigma^0$ the distance is 1.8 $\sigma$.  However, the $\alpha$ 
exponents for K$^0_s$ appears to be lower by 5 $\sigma$ units.
In~\cite{Adamczewski19}, a global value of $\alpha = 1.45 \pm 0.06$ 
was extracted for five strange hadron species emitted from Au+Au collisions 
at the available energy of 2.41 GeV.
This finding is 2.5 $\sigma$ away from our result obtained for all the 
energies, so rather not in disagreement.
We also found that the values of $\alpha$ exponent for the AGS data on 
charged kaon emission at \sqSnn = 4.86 GeV are: $1.18 \pm 0.05$ for K$^+$ 
and $1.22 \pm 0.04$ for K$^-$~\cite{Ahle98}. 
Interestingly, despite the fact that this energy seems to be already 
situated in the region of string degrees of freedom, 
the found values are also within about 2$\sigma$ in agreement with 
the constant average found in our analysis for $\sqSnn \in [2.24,~3.0]$~GeV.

\section{Comparison to transport models}
\label{Sect:transport}

It is of interest to compare the predictive power of the found
parametrization formulae to that of various transport model calculations.
Currently only the collisions of $^{40}$Ar+$^\mathrm{nat}$KCl at the
available energy of 2.61 GeV offer the set of yields for all five 
strange hadrons considered in this work at the same centrality range of 0-35\%. 
We shall treat this data as the basis for comparison. 

The simulations of heavy-ion collisions were performed within the recent
publicly available versions of RQMD.RMF implemented in the Monte-Carlo 
event generator JAM2.1~\cite{RQMD.RMF}, SMASH (\cite{SMASH}, version 2.1) 
and UrQMD (\cite{UrQMD}, version 3.4).
To approximate the $^\mathrm{nat}$KCl, the $^{37}$Ar was taken as the target
nucleus. The class of 35\% most central events was selected by the 
condition of $b < 5.0$~fm, resulting from the Glauber model calculations.
The SMASH simulations were performed for two variants of the Equation 
of State (EoS), characterized by the incompressibility modulus 
$\kappa$ = 240 and 380 MeV. 
To stabilize the mean field map each event was processed in 40 ensembles. 
For RQMD.RMF two momentum-dependent EoS variants were
selected: MD2 and MD4. The variants MD1 and MD3 give results in-between.
As currently none of these variants is pointed out as the more applicable 
one to the considered range of energies, the chosen scenarios can be
treated as an estimate of the systematic error of the model
originating from an uncertainty of EoS. Since we are interested only in the
4$\pi$ yields, the Coulomb potential was neglected both in RQMD.RMF and SMASH.
The publicly available version of UrQMD offers either the cascade approach 
or the "hard" variant of EoS. We chose the latter one due to better
relevance to AA collisions at beam energies in the hadronic sector.

The yields of K$^0_s$ were obtained by dividing the found K$^0$ yield
by 2. The multiplicity of $\phi$ meson was found by counting first the 
so-called ``reconstructable'' K$^+$K$^-$ pairs, i.e. these ones, 
whose invariant mass remains in a peak around 
$m_\phi = 1.0195~\mathrm{GeV}/c^2$. The yield was then corrected 
for BR ($\phi \rightarrow \mathrm{K}^+ \mathrm{K}^-$)~=~49.1~\%~\cite{PDG}.
The yields of $\phi$ mesons from UrQMD 3.4 were not included, 
since the proposed dominant mechanism of $\phi$ production from 
decays of heavy resonances~\cite{Stei16} was introduced after this 
version of the model.
Since the $\Sigma^0$ hyperons were observed in the considered experiments
only indirectly (as the feed-down to the $\Lambda$ spectrum with 
BR $\approx$ 100\%) in our simulations the yields of these two hyperons 
were summed up.

\begin{figure}
  \centering
  \includegraphics[width=0.8\textwidth]{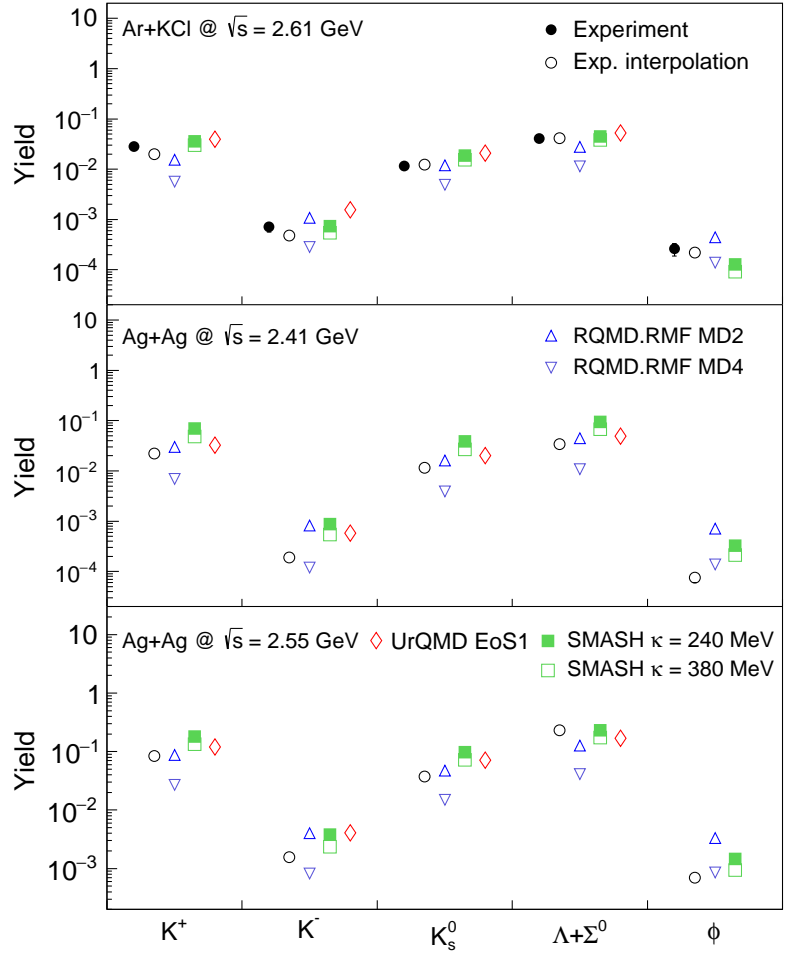}
  \caption{Comparison of experimental, interpolated and simulated 
  production yields of five strange hadrons from
  Ar+KCl at \sqSnn = 2.61 GeV within centrality of \mbox{0-35\%}
  and Ag+Ag at 2.41 and 2.55 GeV within centrality of \mbox{0-10\%}. 
  Data labelled as ``Exp. interpolation'' are the predictions of 
  Eq.~\ref{Eq:fitcurve} using the best-fit parameters from 
  Table~\ref{tab:fitresults}.
  See text for details on the SMASH, RQMD.RMF and UrQMD simulations
  and discussion. 
  }
  \label{fig:comparisonYields}
\end{figure}

In the upper panel of Fig.~\ref{fig:comparisonYields} the yields for
Ar+KCl found experimentally are set together with those obtained from the 
phenomenological parametrization using Eq.~\ref{Eq:fitcurve}, 
and the ones predicted by the transport models.
To quantify the overall discrepancy between the experimental yields
of five considered hadrons and their predictions obtained within a 
given model, we used the total standard deviation:

\begin{equation}
    z_\mathrm{Model} ~=~  \sum_{h} ~ \frac 
    {\left| P_{\mathrm{Exp},h} ~-~  P_{\mathrm{Model},h} \right| }
    {\sqrt{ \Delta P_{\mathrm{Exp},h}^2 ~+~ \Delta P_{\mathrm{Model},h}^2 }}  \quad\quad ,
  \label{eq:TotalDev}
\end{equation}

\noindent where $h$ $\in$ \{K$^+$, K$^-$, K$^0_s$, $\Lambda+\Sigma^0$, $\phi$\}. 
In cases of the transport codes, by the uncertainties of yields 
we mean the statistical errors. For the worst case of the least abundant 
$\phi$ mesons they are at the level of 10\% of the respective yield, 
so their influence is minor compared to the experimental uncertainty.
The values of found $z_\mathrm{Model}$ are listed in Table~\ref{tab:stdevArKCl}.
The phenomenological predictions of Eq.~\ref{Eq:fitcurve}
appear to fare better than RQMD.RMF and UrQMD calculations both overall,
and for each hadron type separately. 
For SMASH, the overall discrepancy is still larger than our prediction. 
However, this model fares better for K$^-$ and in the soft EoS case - for K$^+$.

In two lower panels of Fig.~\ref{fig:comparisonYields} we set together the 
predictions of yields of strange hadrons emitted from 10\% most central 
collisions of Ag+Ag at the available energies of 2.41 and 2.55 GeV. 
These predictions cover both the phenomenological parametrization 
of Eq.~\ref{Eq:fitcurve} and the selected transport models. 
We encourage the researchers to include our parametrization into comparisons 
of the experimental yields of hadrons emitted from these collisions.

\begin{table}[tb]
 \centering
 \begin{tabular}{|c|c|c|c|c|c|c|}
 \hline
Approach                 & K$^+$ & K$^-$ & K$^0_s$ & $\Lambda+\Sigma^0$ & $\phi$ &$\sum h_i$\\
 \hline
Exp. interpolation       &  3.3  &  1.5  &    0.7  &        0.0         &   0.5  &   6.0    \\
 \hline 
 RQMD.RMF MD2            &  5.2  &  2.4  &    0.5  &        4.0         &   2.5  &  14.6    \\
 RQMD.RMF MD4            &  9.3  &  2.9  &    6.6  &        9.0         &   1.7  &  29.4    \\
SMASH $\kappa$~=~240~MeV &  3.3  &  0.2  &    7.0  &        1.2         &   1.8  &  13.5    \\
SMASH $\kappa$~=~380~MeV &  0.8  &  1.1  &    4.0  &        0.8         &   2.3  &   9.0    \\
UrQMD Hard EoS           &  4.6  &  5.6  &    8.5  &        3.1         &   3.6  &  26.7    \\
 \hline
 \end{tabular}
 \caption{Standard deviations (c.f. Eq.~\ref{eq:TotalDev}) 
  between the experimental yields and the ones predicted by different approaches, 
  for hadrons emitted from Ar+KCl at \sqSnn = 2.61 GeV at centrality range of 0-35\%.
  ``Exp. interpolation'' are the predictions of Eq.~\ref{Eq:fitcurve}.
  Last column is the sum of standard deviations over five hadrons.
 }
 \label{tab:stdevArKCl}
\end{table}

\section{Summary}
\label{Sect:summary}

We analysed the available experimental yields of five strange hadron 
types (K$^{\pm,0}$, $\phi$ and $\Lambda$ with feeddown from $\Sigma^0$ decays) 
emitted from collisions of heavy ions at \sqSnn in the range of 2.16 - 3.0 GeV.
Since in the original works the estimation of \Apart was done 
within three different approaches, we used the stated centralities
and extracted the numbers of participants for all the data with 
help of one common method, the Glauber Monte Carlo. 
We then postulated the parametrization of yields of each of 
analysed hadrons as function of \sqSnn and \Apart, and obtained 
the fits with $\chi^2/\nu$ between 0.15 and 3.6. 
We demonstrated the good description of the yields 
at the boundaries of experimental datasets that were used for fitting.
Therefore, we felt encouraged to provide the predictions of yields 
for the currently analysed HADES data from Ag+Ag collisions at the available 
energies of 2.41 and 2.55 GeV, accepted Au+Au experiment at 2.16 and 2.24 GeV, 
some currently unpublished yields from STAR's Au+Au interactions at 3.0 GeV, 
as well as planned CBM data from Au+Au collisions up to 3.85 GeV. 

Overall, regarding the yields, the predictions of our phenomenological
curves for K$^+$ and K$^0_s$ seem to work within [2.16 - 3.0] GeV, 
whereas for K$^-$ the lower limit should be lifted to 2.24 GeV. 
Concerning $\phi$, the range of applicability spans within [2.41 - 3.0] GeV, 
and in case of $\Lambda+\Sigma^0$ the upper limit should be dropped to 2.7 GeV. 
However, in terms of yield ratios, the applicability of K$^+$/K$^0_s$ and 
$\phi$/$K^-$ can be extended up to 3.85 GeV, whereas for K$^-$/K$^+$ 
the fits currently make sense up to 3.32 GeV.

The accumulated data allows us also to find that the $\alpha$ power
exponent of the yield dependence on the \Apart does not change with
available energy within the current uncertainties. 
Its average value was found to be 1.30$\pm$0.02. This value is in
general agreement with the values of $\alpha$ obtained per hadron type
via fit of Eq.~\ref{Eq:fitcurve}, with the exception of K$^0_s$.

We also used the yields from Ar+KCl at 2.61 GeV to benchmark our
parametrization and the predictions of recent publicly available
versions of RQMD.RMF, SMASH and UrQMD. For this system, we concluded
that our approach currently appears to have the best degree of 
overall agreement with the experimental data.
We therefore encourage the researchers to include it into their 
prediction and comparison tools.

\section*{Appendix A}
\label{Sect:appendixA}

\begin{table}[ht]
 \centering
 \begin{tabular}{|c|c|c|c|}
 \hline
  Nucleus & Parametrization &     Parameters  ($R$ [fm], $d$ [fm])              &       Ref.       \\
 \hline
$^{12}$C  &        WF       &                                                   & \cite{Lim19}     \\
 {}       &        FB       &                                                   & \cite{Reuter82}  \\
 {}       &       SoG       &                                                   & \cite{Sick82}    \\
  \hline
$^{27}$Al &       2pF       & $R$ = 3.07, $d$ = 0.519(26)                       & \cite{Vries87}    \\
 {}       &       2pF       & $R$ = 2.84, $d$ = 0.569                           & \cite{Vries87}    \\
 {}       &     2pF-def     & $R$ = 3.07, $d$ = 0.519 , $\beta_2$ = -0.448, $\beta_4$ = 0.239   & \cite{Vries87,Moller95}   \\
 {}       &     2pF-def     & $R$ = 2.84, $d$ = 0.58 , $\beta_2$ = -0.448, $\beta_4$ = 0.239   & \cite{Vries87,Moller95}   \\
 {}       &        FB       &                                                   & \cite{Vries87}    \\
  \hline
$^{36}$Ar &       2pF       & $R$ = 3.53(4)  , $d$ = 0.542(15)                  & \cite{Vries87}    \\
 {}       &       3pF       & $R$ = 3.73(5)  , $d$ = 0.62(1) , $w$ = -0.19(4)   & \cite{Vries87}    \\
 {}       &        FB       &                                                   & \cite{Vries87}    \\
  \hline
$^{39}$K  &       3pF       & $R$ = 3.743(25), $d$ = 0.585(6), $w$ = -0.201(22) & \cite{Vries87}    \\
  \hline
$^{35}$Cl &       3pF       & $R$ = 3.476(32), $d$ = 0.599(5), $w$ = -0.10(2)   & \cite{Vries87}    \\
  \hline
$^{58}$Ni &       3pF       & $R$ = 4.3092(10), $d$ = 0.5169(2), $w$ = -0.1308(10) & \cite{Ficenec70,Vries87} \\
 {}       &        FB       &                                                   & \cite{Vries87}    \\
 {}       &       SoG       &                                                   & \cite{Vries87}    \\
  \hline
$^{63}$Cu &       2pF       & $R$ = 4.163(27), $d$ = 0.606(11)                  & \cite{Vries87}    \\
 {}       &       2pF       & $R$ = 4.218(14), $d$ = 0.596(5)                   & \cite{Vries87}    \\
  \hline
$^{65}$Cu &       2pF       & $R$ = 4.158(35), $d$ = 0.632(8)                   & \cite{Vries87}    \\
 {}       &       2pF       & $R$ = 4.252(15), $d$ = 0.589(5)                   & \cite{Vries87}    \\
  \hline
$^{107}$Ag&       2pF       & $R$ = 5.3006, $d$ = 0.5234                        & \cite{Landolt04}  \\
 {}       &     2pF-def     & as above + $\beta_2$ = 0.162, $\beta_4$ = -0.014  & \cite{Landolt04,Moller95} \\
  \hline
$^{109}$Ag&       2pF       & $R$ = 5.3306, $d$ = 0.5234                        & \cite{Landolt04}  \\
 {}       &     2pF-def     & as above + $\beta_2$ = 0.162, $\beta_4$ = -0.023  & \cite{Landolt04,Moller95} \\
  \hline
$^{197}$Au&       2pF       & $R$ = 6.38(6), $d$ = 0.535(27)                    & \cite{Vries87}    \\
 {}       &     2pF-def     & as above + $\beta_2$ = -0.131, $\beta_4$ = -0.031 & \cite{Vries87,Moller95}  \\
 {}       &        HN       & $R$ = 6.42, $d$ = 0.44                            & \cite{Loizides15} \\
  \hline
 \end{tabular}
 \caption{Parametrization types and values of parameters for density distributions 
 of nuclei used in this work (see text for explanation of symbols).}
 \label{tab:shapes}
\end{table}

In this Appendix we discuss different variants of the nuclear shapes 
that were used for calculations of \Apart and estimation of its 
systematic uncertainty, described in Sect.~\ref{Sect:Apart}.
These variants are shown in Table \ref{tab:shapes}, where 
the following abbreviations are used (see p. 501 in \cite{Vries87}):

\begin{itemize}
 \item 2pF     (two-parameter Fermi form)
 \item 3pF     (three-parameter Fermi form)
 \item FB      (Fourier-Bessel decomposition) 
 \item SoG     (Sum-of-Gaussians decomposition)
 \item 2pF-def (deformed two-parameter Fermi, see Eq. 4 in \cite{Loizides15})
 \item HN      (2pF with recentering method, see Sect. V.C in \cite{Loizides18})
 \item WF      (dedicated wavefunction-based calculations, see \cite{Lim19})
\end{itemize}

\noindent To give predictions on the yields of strange hadrons emitted 
from Ag+Ag collisions, we included also the \Apart values found by our procedure.
For the analysis of the close-to-symmetric Ar + $^\mathrm{nat}$KCl system,
we applied the weighted average the distributions of participants found 
separately for Ar+K and Ar+Cl collisions. 
The weights were the respective collision cross sections obtained from the 
Glauber Monte Carlo simulations.
In case of the natural Cu, we have averaged the results from $A$~=~63 
and $A$~=~65 isotopes with weights corresponding to their contributions.

\section*{Appendix B}
\label{Sect:appendixB}

A stable prediction of the yield of specific hadron at given \sqSnn and \Apart
using Eq.~\ref{Eq:fitcurve}, which includes an extraction of the statistical
error, requires an application of the full covariance matrix up to fourth 
significant digit.
For this reason, we report the relevant matrices in Table~\ref{tab:covmats}. 

\begin{table}[ht]
 \centering
 \begin{tabular}{|c|c|c|c|c|}
  \hline
  K$^+$  &        $N$           &       $\alpha$       &       $\beta$        &         $C$          \\
  \hline
  $N$    & 8.118$\cdot 10^{-7}$ &-5.927$\cdot 10^{-4}$ & 4.177$\cdot 10^{-4}$ & -8.713$\cdot 10^{-6}$ \\
$\alpha$ &        {}            & 4.592$\cdot 10^{-4}$ & 2.514$\cdot 10^{-3}$ & -5.201$\cdot 10^{-5}$ \\
$\beta$  &        {}            &          {}          &       0.2392         & -4.921$\cdot 10^{-3}$ \\
  $C$    &        {}            &          {}          &         {}           &  1.019$\cdot 10^{-4}$ \\
  \hline
  \hline
  K$^-$  &        {}            &          {}          &         {}           &          {}           \\
  \hline
  $N$    & 5.518$\cdot 10^{-9}$ &-2.878$\cdot 10^{-6}$ & 4.900$\cdot 10^{-5}$ & -6.703$\cdot 10^{-7}$ \\
$\alpha$ &        {}            & 1.868$\cdot 10^{-3}$ &-2.185$\cdot 10^{-2}$ &  2.987$\cdot 10^{-4}$ \\
$\beta$  &        {}            &          {}          &       0.4934         & -6.649$\cdot 10^{-3}$ \\
  $C$    &        {}            &          {}          &         {}           &  9.020$\cdot 10^{-5}$ \\
  \hline
  \hline
 K$^0_s$ &        {}            &          {}          &         {}           &          {}           \\
  \hline
  $N$    & 8.853$\cdot 10^{-7}$ &-4.221$\cdot 10^{-5}$ & 6.662$\cdot 10^{-6}$ &          {}           \\
$\alpha$ &        {}            & 2.321$\cdot 10^{-3}$ &-1.973$\cdot 10^{-3}$ &          {}           \\
$\beta$  &        {}            &          {}          & 1.068$\cdot 10^{-2}$ &          {}           \\
  \hline
  \hline
$\Lambda+\Sigma^0$
         &        {}            &          {}          &         {}           &          {}           \\
  \hline
  $N$    & 9.889$\cdot 10^{-9}$ &-4.047$\cdot 10^{-6}$ & 1.020$\cdot 10^{-4}$ &          {}           \\
$\alpha$ &        {}            & 1.724$\cdot 10^{-3}$ &-5.816$\cdot 10^{-2}$ &          {}           \\
$\beta$  &        {}            &          {}          &       41.41          &          {}           \\
  \hline
  \hline
$\phi$
         &        {}            &          {}          &         {}           &          {}           \\
  \hline
  $N$    & 2.067$\cdot 10^{-10}$&-1.681$\cdot 10^{-6}$ & 4.431$\cdot 10^{-7}$ &          {}           \\
$\alpha$ &        {}            & 1.426$\cdot 10^{-2}$ &-7.206$\cdot 10^{-3}$ &          {}           \\
$\beta$  &        {}            &          {}          & 5.268$\cdot 10^{-2}$ &          {}           \\
  \hline
 \end{tabular}
 \caption{Covariance matrices of the best fits of Eq.~\ref{Eq:fitcurve} 
   to experimental yields of studied strange hadrons.
   All the matrices were found to be symmetric.
   For K$^0_s$, $\Lambda + \Sigma^0$ and $\phi$ the $C$ parameter 
   was fixed during fitting.}
 \label{tab:covmats}
\end{table}

\section*{Acknowledgement}

We thank Tomasz Matulewicz for valuable discussions and comments to the manuscript.

\end{document}